\documentclass[superscriptaddress,pre]{revtex4}
\usepackage{epsfig,amsmath,amssymb,graphics,color,calc,psfrag}

\newcommand{\be}{\begin{equation}}
\newcommand{\ee}{\end{equation}}
\newcommand{\ba}{\begin{eqnarray}}
\newcommand{\ea}{\end{eqnarray}}

\begin{document}

\title{What do we learn from the shape of 
the dynamical susceptibility of glass-formers?}

\author{Cristina Toninelli}
\affiliation{ENS 24 rue Lhomond, 75231 Paris Cedex 05, France.}

\author{Matthieu Wyart}
\affiliation{Service de Physique de l'{\'E}tat Condens{\'e}
Orme des Merisiers --- CEA Saclay, 91191 Gif sur Yvette Cedex, France.}

\author{Ludovic Berthier}
\affiliation{Laboratoire des Verres UMR 5587, Universit\'e Montpellier
II and CNRS, 34095 Montpellier, France}

\author{Giulio Biroli}
\affiliation{Service de Physique Th{{\'e}o}rique
Orme des Merisiers --- CEA Saclay, 91191 Gif sur Yvette Cedex, France.}

\author{Jean-Philippe Bouchaud}
\affiliation{Service de Physique de l'{\'E}tat Condens{\'e}
Orme des Merisiers --- CEA Saclay, 91191 Gif sur Yvette Cedex, France.}
\affiliation{Science \& Finance, Capital Fund Management
6-8 Bd Haussmann, 75009 Paris, France.}

\date{\today}

\begin{abstract}
We compute analytically and numerically
the four-point correlation function that characterizes
non-trivial cooperative dynamics in glassy systems within 
several models of glasses: elasto-plastic deformations, 
mode-coupling theory (MCT), collectively rearranging regions (CRR), diffusing
defects and kinetically constrained models (KCM). 
Some features of the four-point susceptibility $\chi_4(t)$ are expected to be universal: 
at short times we expect a power-law increase in time as $t^{4}$ due to ballistic motion 
($t^{2}$ if the dynamics is Brownian) followed by an elastic regime (most relevant deep in the 
glass phase) characterized by a $t$ or $\sqrt{t}$ growth,
depending on whether phonons are propagative or diffusive. We find both in the 
$\beta $, and the early $\alpha $ 
regime that $\chi_4 \sim t^\mu$, where $\mu$ is 
directly related to the mechanism responsible for relaxation. This regime ends when
a maximum of $\chi_4$ is reached at a time $t=t^*$ of the order of
the relaxation time of the system. This maximum is followed by a fast decay to zero at large times.
The height of the maximum also follows a power-law, $\chi_4(t^*) \sim t^{*\lambda}$.
The value of the exponents $\mu$ and $\lambda$ allows one 
to distinguish between different mechanisms. 
For example, freely diffusing defects in $d=3$ lead to 
$\mu=2$ and $\lambda=1$, whereas the CRR
scenario rather predicts either $\mu=1$ or a logarithmic behaviour depending on the nature of
the nucleation events, and a logarithmic behaviour of $\chi_4(t^*)$. MCT leads to $\mu=b$ and $\lambda
=1/\gamma $, where $b$ and $\gamma$ are the standard MCT exponents. 
We compare our theoretical results with numerical simulations on a Lennard-Jones and a
soft-sphere system. Within the limited time-scales accessible to numerical simulations, 
we find that the exponent $\mu$ is rather small, $\mu < 1$, with a value in
reasonable agreement with the MCT predictions, but not with 
the prediction of simple diffusive defect models, KCMs with non-cooperative defects and CRR. 
Experimental and numerical determination of $\chi_4(t)$ for longer time scales
and lower temperatures would yield highly valuable
information on the glass formation mechanism. 
\end{abstract}

\maketitle

\section{Introduction} 

The idea that the sharp slowing down of super-cooled liquids is related to
the growth of a cooperative length scale dates back at least to 
Adam and Gibbs \cite{AG}. 
But it is only a few years back that this idea has started being substantiated 
by convincing experiments \cite{Ediger,Richert,Weeks,Israeloff,Vandenbout}, 
numerical simulations \cite{Harrowell,Harrowell2,Onuki,parisi,heuer,Glotzer,Glotzer2,hiwatari} 
and simple microscopic 
models~\cite{SR,Jaeckle,Sollich,nef,FA,GC,WBG,WBG2,BG,KA1,TBF,pan}.
One of the basic problem has been to find an observable that allows one to 
define and measure objectively such a cooperative length scale. An interesting 
quantity, proposed a few years ago in the context of mean field $p$-spin
glasses \cite{FP} (see \cite{KT} for an important early insight) 
and measured in simulations, is a four-point density correlator, defined as
\begin{equation}
G_4(\vec r,t) = \langle \rho(0,0)  \rho(0,t) 
\rho(\vec r,0) 
\rho(\vec r,t) \rangle-\langle \rho(0,0)  \rho(0,t)  \rangle \langle
\rho(\vec r,0) 
\rho(\vec r,t) \rangle,\label{G4}
\end{equation}
where $\rho (\vec r,t)$ represents the density fluctuations at
position $\vec r$ and time $t$. In
practice one has to introduce an overlap function $w$~\cite{FP},
to avoid singularity due to the evaluation of the density at the same
point or consider slightly different correlation functions \cite{Berthier2}.
This quantity measures the correlation in space of local time correlation
functions. Intuitively if at point $0$ an event has occurred that 
leads to a decorrelation of the local density over the time 
scale $t$, 
$G_4(\vec r,t)$ measures the probability that a similar event has 
occurred a distance $\vec r$ away 
within the same time interval $t$ (see e.g.~\cite{mayer}). 
Therefore $G_4(\vec r,t)$ is a 
candidate to measure heterogeneity and 
cooperativity of the dynamics. The best theoretical justification for 
studying this quantity is to
realize that the order parameter for the glass transition is already a 
two-body object, namely 
the density-density correlation 
function $C(t)=\langle \rho(0,0)  \rho(0,t) \rangle$, which decays to 
zero in the liquid phase and
to a constant value in the frozen phase. The four-point correlation 
$G_4(\vec r,t)$ therefore 
plays the same role as the standard two-point correlation function 
for a one-body order parameter 
in usual phase transitions. Correspondingly, the associated susceptibility 
$\chi_4(t)$ is defined as the volume integral of $G_4(\vec r,t)$, 
and is equal to the
variance of the correlation function \cite{FP,Berthier1,BB1}. The
susceptibility $\chi_4(t)$
has been computed numerically for different model glass formers, 
and indeed exhibits a maximum 
for $t = t^* \sim \tau_\alpha$, the relaxation time of the system
\cite{heuer,Glotzer,Glotzer2,hiwatari}. The peak value  
$\chi_4(t^*)$ is seen to increase as the temperature decreases, 
indicative that the range of 
$G_4(\vec r,t^*)$ increases as the system becomes more sluggish.
The dynamical correlation length $\xi_4 (t^*)$ extracted from $G_4(\vec
r,t^*)$ in molecular dynamics simulations grows and becomes
of the order of roughly $10$ inter-particle distances when the
time-scales is of the order of $10^{5}$ microscopic time-scales $\tau_0$
with $\tau_0 \sim 0.1$~ps for an atomic liquid.
In experiments close to the glass transition the dynamical correlation length
has been found to be only slightly larger, between $10$ and $20$
inter-particle distances \cite{Ediger,Richert}. This is puzzling because 
experiments are done on systems with relaxation times
that are several orders of magnitude larger than in simulations. 
In fact, extrapolating simulation results in the experimental regime
would lead to much larger dynamical correlation lengths.
The origin of this puzzle is still unclear, see Ref.~\cite{nef} for a recent discussion.
Experiments on dynamical heterogeneity bridging the gap 
between numerical and macroscopic time-scales would be extremely valuable to resolve this paradox.

Several scenarii have been proposed to understand the existence of 
non-trivial dynamical correlations, and their relation with thermodynamical
singularities. Adam and Gibbs \cite{AG}, Kirkpatrick, Thirumalai, Wolynes and collaborators~\cite{KTW} 
(for a different formulation, see Ref.~\cite{BB2}), and 
Kivelson and Tarjus~\cite{Tarjus} have proposed, using somewhat different arguments, the idea of collectively rearranging 
regions (CRR), of size $\xi$ that increases as the temperature is decreased. The evolution of the system is such
that these regions are either frozen or allowed to temporarily and collectively unjam for a short time 
until a new jammed configuration is found. 

In apparent contradiction with the existence of the growing length scale, the 
Mode Coupling Theory (MCT) of glasses states that the self consistent
freezing of particles in their cages is a purely local
process with no diverging length scale at the transition
\cite{Gotze}. However, this point of view is in disagreement with the results found
for mean-field disordered systems \cite{KT,FP} that are conjectured to
provide a mean-field description of the glass transition and 
display an MCT-like dynamical transition. Indeed it was recently 
shown that within MCT 
$G_4(\vec r,t)$ in fact develops long range correlations close to the 
critical MCT temperature $T_c$ \cite{BB1}. Within a phase-space
interpretation of the MCT transition, the mechanism 
for this cooperative behaviour for $T > T_c$ is the progressive 
rarefaction of energy lowering directions~\cite{Grigera}.
Within a real-space interpretation, the MCT transition is due to the formation of a large number of
metastable states, each one characterized by a surface
tension that increases from zero at $T_c$. As one approaches $T_c$ from above, 
the relevant eigenvectors of the dynamical Hessian become more and more extended, which 
means that
the modes of motion that allow the system to decorrelate are made of very well defined, 
collective rearrangements of larger and larger clusters of particles (see the recent work
of Montanari and Semerjian \cite{MS}).
For smaller temperatures, $T < T_c$, `activated events' 
are expected to play a crucial role. They 
are believed to be responsible for the destruction of the freezing
transition at $T_c$. This regime has been tentatively described by
adding `hopping terms' in the MCT equations \cite{Gotze} or 
within a CRR scenario \cite{KTW,BB2}.

Exploiting yet a different set of ideas, models of dynamical facilitation,
such as the Frederickson-Andersen~\cite{FA} or
Kob-Andersen models~\cite{KA1}, 
have recently been proposed as paradigms for glassy 
dynamics~\cite{SR,GC,BG}.
In these models, 
the motion of particles is triggered by `mobility defects' that
diffuse and possibly interact within the system. As the 
temperature is lowered or the density is increased, the concentration of 
defects goes down, and the relaxation
time of the system increases. Dynamics is obviously heterogeneous 
since it is catalyzed by defects that cannot be everywhere simultaneously. 
The characteristic length scale in this case is 
related to the average distance between defects to some model and 
dimension dependent exponent~\cite{SR,GC,BG,TBF,nef}.

Understanding the mechanism behind the growth of the dynamical
correlation length is certainly an important step---arguably the most
important one---to understand the cause of the slowing down 
of the dynamics.
Furthermore, the different scenarii for the glass transition can be tested 
contrasting their quantitative prediction for the four-point 
correlation function $G_4(\vec
r,t)$ to the numerical, and hopefully soon experimental, results.
Following these premises we investigate in this paper
the analytical shape of $G_4(\vec r,t)$ for several simple models. 
We show that $G_4(\vec r,t)$ indeed contains some important
information concerning the basic relaxation mechanisms. However we show 
that, perhaps disappointingly,
models where cooperativity is absent or trivial lead to four-point 
correlation functions and dynamical 
susceptibilities $\chi_4$ that exhibit non trivial features. Other, 
more complex observables will have to be defined to
really grasp the nature of the collective motions involved in the 
relaxation process of glasses~\cite{HeuerStrings,Harrowell2}.

Let us summarize the main results of our study, in terms of the
susceptibility $\chi_4(t)$ and time-sectors. In a super-cooled liquid
there are separate regimes of time-scales corresponding to different physical
behaviour (see Fig. 1). On microscopic time-scales 
particles move ballistically if the dynamics is Newtonian, or 
diffusively if the dynamics is Brownian. On a longer time-scale, interactions start playing
a role, which can be described approximately using elasticity theory, before a truly 
collective phenomenon sets in. This non trivial glassy regime is the $\beta$-regime, 
within which correlation
functions, as for example the dynamical structure factor, develop a
plateau. The $\beta$-regime is divided further
in an early and a late $\beta$-regime corresponding, respectively, to the 
approach and the departure from the plateau of the correlation function. 
Finally the structural relaxation
time-scale on which correlation functions decay to zero is
the $\alpha$-regime. All previous studies have focused on the behaviour of $\chi_4(t)$ at
times of the order of $\tau_{\alpha }$ which correspond to the peak of $\chi_4(t)$.
We show that $\chi_4(t)$ has in fact a rich structure in time and
different behaviour in different time-sectors. In many of these regimes, $\chi_4(t)$ 
behaves as a power-law of time, $t^\mu$, with different values of $\mu$.
During the ballistic time-scale one finds $\mu=4$ 
($\mu={2}$ for Brownian dynamics) whereas during the elastic regime (most relevant deep in the
glass phase), the exponent becomes $\mu=1$ for ballistic phonons and 
$\mu=1/2$ for diffusive phonons.
The behaviour in the $\beta$ and $\alpha$ regimes is intimately
related to the physical mechanism for relaxation and indeed we find
quite different answers depending on which scenario we focus on.
MCT predicts exponents $\mu={a}$ and $\mu={b}$ on time-scales
corresponding respectively to the early and late $\beta$ regimes, 
where $a$ and $b$ are the standard MCT exponents obtained from the study of the 
dynamical structure factor. The power-law $t^{b}$ extends until the
peak in $\chi_4(t)$ is reached. 

\begin{figure}
\psfig{file=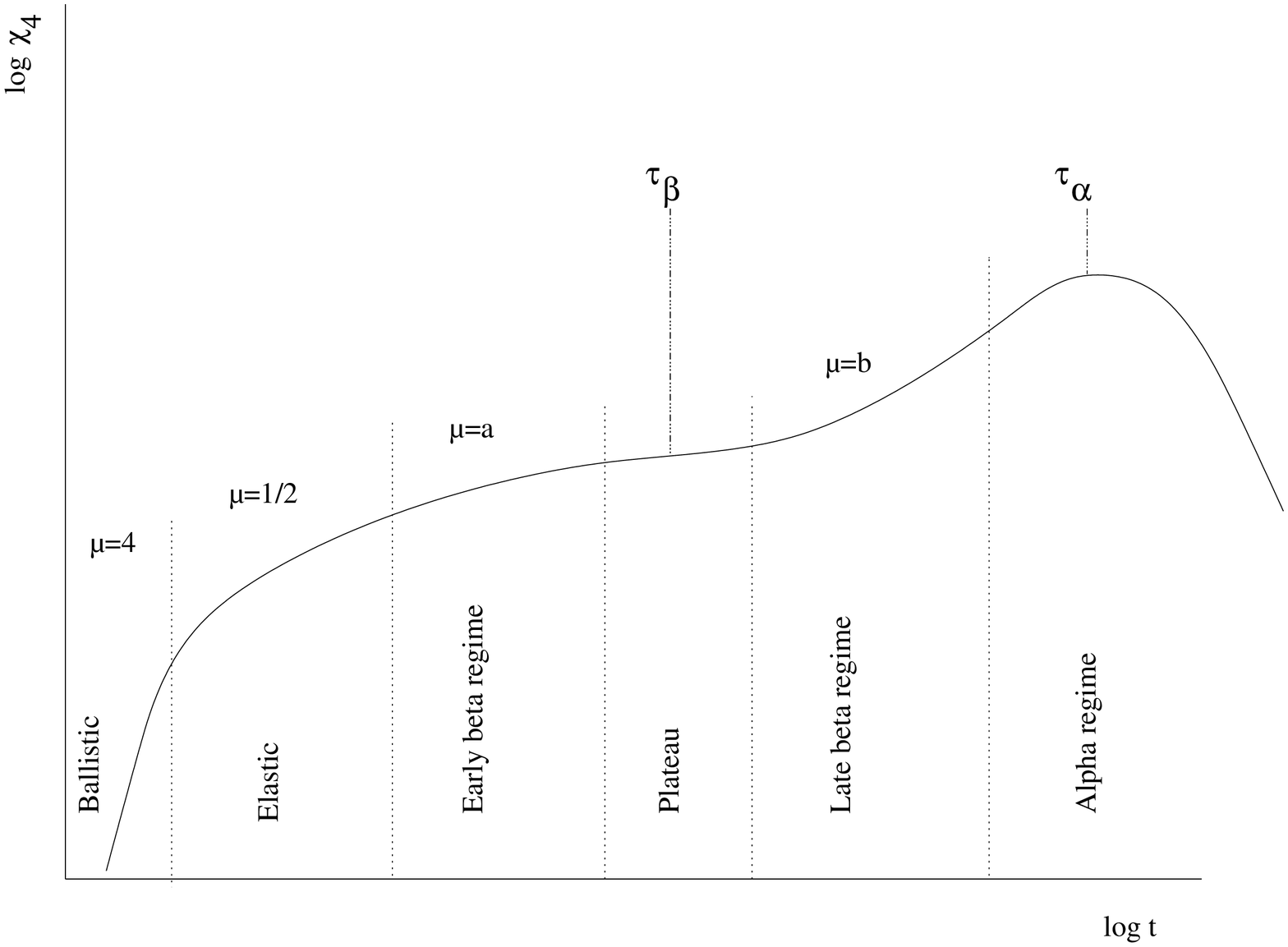,width=9.5cm}
\caption{\label{mctfig} 
Sketch of the time behaviour of $\chi_4(t)$, with all the different time regimes, 
within the MCT description that we find to be a good description around $T_c$. As the
temperature is lowered, we expect the elastic regime to extend up to $\tau_\alpha$.} 
\end{figure}

The other scenarii only make predictions in the $\alpha$ regime. 
In the case of CRR one has $\chi_4 \sim t$ or $\chi_4 \sim (\ln  t)^{d+1/\psi }$ before the peak
depending whether one assumes that the relaxation occurs via bulk nucleation events 
or domain wall fluctuations, see below. For diffusing 
defects in dimension $d=3$, the exponent is $\mu={2}$. If defects have a non trivial 
diffusion exponent $z$, such that their displacement 
at time $t$ scales as $t^{1/z}$, then $\mu=2d/z$ for $d < z$ and
$\mu=2$ otherwise. The overall behaviour of $\chi_4(t)$ is summarized by Fig.~\ref{mctfig}, which specializes to
the MCT predictions for simplicity. 

Another important feature of $\chi_4$ is the growth of the
peak compared to the growth of the time $t=t^* \sim \tau_\alpha$ at
which the peak takes places~\cite{WBG}. This is found to scale as
$\chi_4(t^*) \sim t^{*\lambda}$,
with $\lambda=0$ (logarithm) for CRR, $\lambda=1$ for freely diffusing
defects, $\lambda=d/z$ for anomalously diffusing
defects for $d < z$ and $\lambda=1$ again for $d > z$. 
Note that if the defect diffusion coefficient itself scales with
$t^*$ as $1/t^{*f}$, as for example in the one-spin
facilitated FA model, there is an extra contribution that gives $\lambda=1-f$ for $d > z$. 
Finally, one has $\lambda=1/\gamma$ in the context of MCT, where 
$\gamma$ describes the power-law divergence of the relaxation time as 
the critical MCT temperature is approached. 

We have checked these predictions in two model systems of glass-forming 
liquids: 
a Lennard-Jones and a soft-sphere mixture. Concerning the behaviour of 
$\chi_4(t)$ in the late $\beta$ and $\alpha$ 
regime, the most interesting time-sectors, we have found reasonable agreement with 
the MCT predictions for four point correlators. This agreement is by no means trivial
and is actually quite unexpected unless MCT indeed captures some of the physics of the problem. 
Instead models of diffusing defects do not describe 
well the numerical results unless one assumes
anomalously diffusing defects with $z$ substantially smaller than $2$.
This is perhaps not very surprising since we are focusing on two fragile 
liquids (at least in the numerical time window) at temperatures 
well above the experimental glass transition. It might be 
that the predictions of these models work only on larger time-scales.
In any case, we expect instead that for strong liquids displaying 
an Arrhenius behaviour the predictions for $\chi_4(t)$ obtained studying model of simple diffusing 
defects should hold quantitatively, since
it is indeed quite well established that relaxation in strong liquids 
is triggered by the diffusion of connectivity defects~\cite{Angell,OldMDpapersSIO2}. 
Finally, the CRR picture does not agree quantitatively with our present numerical data. 
However, this picture is supposed to describe the liquid dynamics precisely in the low temperature/
long time regime that is presently beyond numerical capabilities. 
Again, experimental results probing the 
behaviour of $\chi_4(t)$ in this regime would be highly valuable to put strong constraints on the 
different theoretical scenarii of glass formation.

The organization of the paper is as follows. In
Section~\ref{Short-time behaviour} we discuss the behaviour 
of $\chi_4(t)$ on microscopic time-scales. Then, we analyze the predictions of 
elasticity theory in Section~\ref{Elastic}. In Sections~\ref{MCT} and \ref{CRR}
we focus on the behaviour of $\chi_4(t)$ in the $\beta$ and $\alpha$
regimes for MCT and CRR. In Section~\ref{defect} we discuss the
predictions of defect models analytically using an independent 
defect approximation and by numerical simulations of 
kinetically constrained models. 
In Section \ref{numerics} we compare the different predictions to the
results of numerical simulations of models of glass-forming liquids.
We present our conclusions in Section \ref{conclusion}.

\section{Microscopic dynamics}
\label{Short-time behaviour}

On very short time-scales the behaviour of $\chi_4$ can be computed exactly.  
For simplicity, we characterize the  dynamics through the 
self-intermediate scattering function, 
\begin{equation}
F_s(k,t) = \frac{1}{N} \sum_i \left\langle
\cos { \vec{k} \cdot [\vec{r}_i(t) - \vec{r}_i(0) ]} \right\rangle ,
\end{equation}
and define the dynamic susceptibility as the variance 
of the fluctuations of $F_{s} (k,t)$: 
\begin{equation}
\chi_4(t)=N\left[ \left\langle  \left( \frac{1}{N} \sum_i 
\cos { \vec{k} \cdot [\vec{r}_i(t) - \vec{r}_i(0) ]}\right)^2
 \right\rangle-\left\langle \frac{1}{N}  \sum_i 
\cos { \vec{k} \cdot [\vec{r}_i(t) - \vec{r}_i(0) ]}
 \right\rangle^{2}\right].
\end{equation}
The full intermediate four point scattering function defined in Eq.~(\ref{G4}) 
in fact contains very similar information, even for interacting systems -- as shown
by numerical simulations \cite{Glotzer,Glotzer2}.

On a very short time-scale particles move ballistically if the dynamics
is Newtonian, $\vec{r}_i(t) - \vec{r}_i(0)=\vec{v}_{i}t+O(t^{2})$, 
where $\vec{v}_{i}$ is
the velocity of the particle $i$ at time $t$. Since the system is in
equilibrium all the $\vec{v}_{i}$'s are independent Gaussian variables
with variance $\langle \vec{v}_{i}\cdot \vec{v}_{j} \rangle
=\delta_{ij}3k_{B}T/m$, 
where $T$ is
the temperature, $m$ the mass of the particles, 
and $k_{B}$ the Boltzmann
constant. Using this property it is straightforward to obtain
\begin{equation}
F_s(k,t) = \exp \left(-\vec{k}^{2}\frac{k_{B}T}{2m}t^{2} \right)
\end{equation}
and
\be \label{e1}
\chi_4 (t)=
F_s(k,t)^{2}\left[\cosh \left(-2\vec{k}^{2}\frac{k_{B}T}{m}t^{2} \right)-
1\right]
\ee
For an interacting particle systems this is only valid on 
short time scales, for example smaller than the
collision time for short ranged interactions.  
This leads to an initial power-law increase that reads
\begin{equation}\label{e2}
\chi_4 (t)=
\frac{1}{2} (\vec{k}^{2})^{2}\left(\frac{k_{B}T}{m}\right)^{2}t^{4}+O (t^{6}).
\end{equation}
Note that if one had chosen Langevin dynamics (i.e. 
$\partial_{t}\vec{r}_{i}=\partial_{\vec{r}}H+\vec{\eta}_{i}$) instead
of Newtonian dynamics, Eqs. (\ref{e1},\ref{e2}) would have been
identical except for the replacement of
$k_{B} T t^{2}/m$ by $2Tt$, again for small times. Thus changing from Newtonian
to Langevin dynamics, the initial power-law increase of $\chi_4 (t)$
changes from $t^{4}$ to $t^{2}$. 
This is similar to the change in the mean square
displacement that increases as $t^{2}$ and $t$, respectively, for
Newtonian and Langevin dynamics.

In the above example, however, it is clear that the increase of $\chi_4$ with time has nothing to do with 
the increase of a correlation length, since particles are assumed to be independent. In other words, the 
four-point correlation $G_4(\vec r,t)$ has a trivial $\delta$-function spatial dependence, but the height of
the $\delta$ peak increases with time. As will be discussed later in the paper, it is important to normalize 
$\chi_4(t)$ by the value of $G_4(\vec r=0,t)$ to conclude from the four-point susceptibility that a length scale is 
indeed growing in the system.  

\section{Elastic contribution}
\label{Elastic}

For longer time-scales the interaction between particles starts playing a r\^ole. Generically 
one expects that in the time regime where the displacements of particles remain small, an elastic description
should be valid. In a solid, or in a glass deep below $T_g$, there is no further relaxation channels and the
elastic contribution to $\chi_4$ should be the only relevant one. In a super-cooled liquid around the Mode-Coupling temperature $T_c$, the elastic regime is interrupted by
the collective $\beta$ regime, where in some sense phonon-phonon 
interactions completely change the physical picture. Although we expect such 
a crossover, we have at present no detailed theoretical description of it.  

In the following we analyze again the behaviour of the four point self-intermediate scattering function 
assuming that the dynamical behaviour of the liquid can be described, within a restricted time-sector, 
as an elastic network (we will discuss later how to include, in a phenomenological way, viscous flow).
Perhaps surprisingly, we find a non trivial structure for $G_4$ in this model, with an ever growing 
`cooperative' length scale which comes from the dynamics of phonons, that represent the simplest form
of cooperativity. 

We consider an isotropic solid immersed in a viscous thermal bath. The energy of the system is given by:
\be
H=\int d^dr \frac{1}{2} \kappa_1 [\sum_i u_{ii}]^2 + \kappa_2 \sum_{i,j}u_{i,j}^2
\ee
where $\kappa_1,\kappa_2$ are the Lam{\'e} coefficient, 
$u_{i,j}=\frac{1}{2}[\frac{d\phi_i}{dx_j}+\frac{d\phi_j}{dx_i}]$ is the deformation tensor and 
$\vec{\phi}$ the displacement field from an undeformed reference state. Note that $\vec{\phi}(x)$ is 
simply the continuum limit of the displacement of each particle with respect to its equilibrium (bottom of the well)
position.

As is well known, the above 
energy leads to three independent phonon modes (one longitudinal and two transverse modes). For simplicity, we
only consider one deformation mode and write the Hamiltonian in Fourier space as:
\be
H= \frac12 \kappa \int \frac{d^dk}{(2\pi)^d}  k^2 \phi_k \phi_{-k},
\ee
where $\kappa$ is an effective elasticity modulus. The mode $k$ has an energy 
$E_k= \kappa k^2 \, \phi_k\phi_{-k}/2$ and therefore we expect, in equilibrium, $\langle \phi_k\phi_{-k}\rangle
= T/\kappa k^2$, where the Boltzmann constant has been set to unity. Our goal is to calculate the dynamical 
correlation functions of the system. We describe the dynamics by a Langevin equation with a local noise:
\be
m \frac{\partial^2 \phi(\vec r,t)}{\partial t^2} + \nu \frac{\partial \phi(\vec r,t)}{\partial t}= 
\kappa \Delta \phi(\vec r,t)+ \zeta(\vec r,t),
\ee
where $\zeta(x,t)$ is a Gaussian noise uncorrelated in space and time, of variance equal to $2\nu T$. 
Taking the Fourier transform:
\be
m \frac{\partial^2 \phi_k}{\partial t^2} + \nu \frac{\partial \phi_k}{\partial t}= -\kappa k^2 \phi_k + \zeta_k(t)
\ee
$\zeta_k(t)$ is again a Gaussian noise uncorrelated for different $k$'s and time.

In this section, we only consider in details the over-damped case $m=0$ and set $D=\kappa/\nu$, but also give at 
the end the result for the purely propagative case $\nu=0$ (see also Appendix A). One easily deduces the non 
equal time correlation in the over-damped case:
\be
\langle \phi_k(t)\phi_{-k}(0) \rangle=\frac{T}{\kappa k^2} e^{-D k^2 t}. 
\ee

Let us now define the function:
\be
F^{(q)}(r,t)=\sum_i \delta(r-r_i(0))\cos (q[r_i(t)-r_i(0)]),
\ee
whose average equals the self-intermediate scattering function up to a 
constant (the particle density).

Using the microscopic definition of $\vec{\phi}$ we obtain that
\be
 C(q,t)=\langle F^{(q)}(r,t)\rangle\simeq \langle e^{iq[\phi(\vec r,t)-\phi(\vec r,0)]}\rangle =
e^{-q^2\langle [(\phi(\vec r,t)-\phi(\vec r,0)]^2\rangle/2},
\ee
where the last equality comes from the Gaussian nature of the deformation field. Using the above results on the 
correlation of the Fourier modes, we find:
\be
\langle [\phi(\vec r,t)-\phi(\vec r,0)]^2 \rangle= \frac{2T}{\kappa} 
\int\frac{1-e^{-D k^2 t}}{k^2}\, \frac{d^dk}{(2\pi)^d}
\ee
As is well known, this integral behaves differently for $d \le 2$ and for $d > 2$, reflecting the fact that
phonons destroy translational order in low dimensions. As above, we will only consider here the physical 
case $d=3$, relegating the discussion of the other cases to Appendix A. For $d=3$, we need to introduce an
ultraviolet cutoff $\Lambda$ on the wavevector $k$, which is the inverse of the underlying lattice spacing $a$. 
Then, the above integral goes to a constant $\propto \Lambda$ at large times, reflecting the fact that 
particles are localized in their `cage'. Therefore, the self-intermediate scattering function
$C(q,t)$ decays at small times $\Lambda^2 Dt \ll 1$ before saturating to a `plateau' value given by:
\be
f_q \equiv C(q, t \to \infty) = \exp\left(-c \frac{T \Lambda q^2}{\kappa}\right),
\ee
where $c$ is a numerical constant. [Note that $T \Lambda q^2/{\kappa}$ has no dimension, and is expected, from a 
Lindemann criterion, to be of the order of $0.05$ 
at half the melting temperature and for $q = \Lambda$]. 
In real glass-forming liquids, this plateau phase does not persist for ever, 
and $C(q,t)$ finally decays to zero beyond $t=\tau_\alpha$, in the so-called $\alpha$-relaxation regime. 
A modification of the model to account for this decorrelation will be discussed later. Furthermore, the above 
pseudo-$\beta$ regime predicted by elastic theory does not explain 
quantitatively the $\beta$-regime in
super-cooled fragile liquids, except probably on relatively short time scales, 
say up to a few picoseconds. On the other hand, at temperatures below $T_g$ or
for strong glasses, we expect that the elastic regime will extend up to 
$\tau_\alpha$ and compete with other mechanisms, such as the defect mediated 
correlation discussed in Section~\ref{defect} below.  

The calculation of $G_4^{(q)}(\vec r,t)=\langle F^{(q)}(r',t)F^{(q)}(r'+r,t)\rangle _c $ is detailed in Appendix A. 
One immediately sees that $G_4^{(q)}(\vec r,t)$ is governed 
by a diffusive correlation length  $\xi(t) \sim \sqrt{D t}$ with $D=\kappa/\nu$, as
expected from the structure of the Langevin equation that describes
relaxational dynamics. Clearly, in the case of propagative phonons, one finds $\xi(t) \sim Vt$ with $V^2=\kappa/m$. 
The final result, see Appendix A, is:
\be \label{eqel}
G_4^{(q)}(\vec r,t)=C^2(q,t)\left(\cosh(2q^2 R(\vec r,t))-1\right),
\ee
where 
\be
R(\vec r,t)=\frac{T}{\kappa} (Dt)^{1-d/2} F\left(\frac{r}{\sqrt{D t}}\right)
\ee
and we find (see Appendix A)  $F(z) \simeq (4\pi z)^{-1}$ for $z \ll
1$ and $F(z) \simeq (2 \pi^{3/2})^{-1} \exp(-z^2/4)/z^2$ for $z \gg 1$.
Note the similarity between the expression in (\ref{eqel}) and the
corresponding one (\ref{e1}) derived in the previous section. One can
check that indeed the short time behaviour is indeed the one
derived before in the case of Langevin dynamics for the particles, as expected.
Let us now focus on long-times, but still within the elastic regime: $\Lambda^2 Dt \gg 1$,
and for $r \ll \xi(t)$:
\be \label{eqg4el}
G_4^{(q)}(\vec r,t)=f_q^2 \left(\cosh(\frac{T q^2}{2\pi \kappa r})-1\right).
\ee
Suppose for simplicity that we are in a regime where the argument of the $\cosh$ is always small, corresponding 
to the limit $T q^2 \Lambda \ll \kappa$ (remember that by definition $r > a = 2\pi/\Lambda$, where $a$ is the
inter-atomic distance). Then, $G_4(\vec r,t) \sim r^{-2}$
for $\Lambda^{-1} \ll r \ll \xi(t)$. For larger scales $r \gg \xi(t)$ decays as a Gaussian, i.e. 
super-exponentially fast. Note that the small $r$ behaviour of $G_4(\vec r,t)$ is not of the Ornstein-Zernike form 
($1/r$ in $d=3$). Integrating $G_4$ over $\vec r$ we find the dynamical susceptibility,  
\be\label{chi4elas}
\chi_4^{(q)}(t) \sim \frac{T^2 q^4 f_q^2}{\kappa^2} \xi(t).
\ee
This result is actually valid both for in the diffusive limit where $\xi(t) = \sqrt{Dt}$ and in the 
propagative regime where $\xi(t)=Vt$. Therefore $\chi_4^{(q)}(t)$ increases either as $\sqrt{t}$ 
or as $t$ (note that in the limit of small times one recovers the $t^4$ or $t^2$ laws obtained in 
the previous section). In the general case, one expects a crossover between a propagative regime at small times
$t < m/\nu = D/V^2$ (of the order of ps in glass formers, see \cite{pico}) and 
a diffusive regime for longer time scales. Thus, looking at $\chi_4^{(q)}(t)$ as a function of time in a 
log-log plot one should see first a straight 
line corresponding to the ballistic or diffusive motion 
leading respectively to slope $\mu=4$ or $\mu=2$, bending over towards a smaller 
slope ($1$ or $1/2$, or both depending on the strength of the dissipation). The order of magnitude of 
$\chi_4^{(q)}(t)$, as given by Eq.~(\ref{chi4elas}), can be estimated to be $\sim 10^{-3}-10^{-2} 
a^2 \xi(t)$ for
$q= \Lambda$. In the propagative regime with $t = 1$ ps, 
$V=3\, 10^3$ m/s, $a = 0.3$ nm, one finds
$\xi = 10 a$ and $\chi_4^{(q)} \sim 10^{-2}-10^{-1} a^3$, i.e. a small, but perhaps detectable 
signal from the phonons. Only on much larger time scales will the
elastic contribution be significant, a regime that can be reached deep in the glass phase
\cite{foot}.
As mentioned above, 
other collective modes come into play in super-cooled fragile liquids, 
in particular around the mode-coupling temperature, 
and give rise to the $\beta$-regime where `cages' themselves 
become more complex, extended objects \cite{BB1}.

The above calculation shows that in an elastic solid with diffusive or propagative phonon modes, 
the dynamical susceptibility increases without bound, reflecting the presence of Goldstone soft-modes in the system.  
Of course, in a real glass the correlation function $C^{(q)}(t)$ 
eventually decays to zero beyond the $\alpha$ relaxation time $\tau_\alpha$, as particles start 
diffusing out of their cages, far away from their initial position. If phonons were the only 
relevant excitations, this would cause the dynamical susceptibility to peak around $t=t^*=\tau_\alpha$.
A phenomenological model that describes the decay of $\chi_4^{(q)}(t)$ within the above 
elastic framework is to assume a (Maxwell) viscoelastic local modulus:
\be
\frac{\partial \phi(\vec r,t)}{\partial t}= 
\kappa \left[\int_{-\infty}^t dt' 
e^{-\gamma(t-t')} \Delta \frac{\partial\phi(\vec r,t')}{\partial t'}\right]+ 
\zeta(\vec r,t),
\ee
with $\gamma \sim \tau_\alpha^{-1}$, corresponding to a frequency dependent elastic modulus 
$\kappa(\omega)=i\kappa \omega/(i \omega + \gamma)$.
In this model, the dynamics of $\phi$ becomes diffusive at times $> \gamma^{-1}$, and the dynamic structure 
factor therefore decays exponentially beyond that time. Of course, the model itself becomes inconsistent 
at large times, since the underlying lattice needed to define the deformation field $\phi$ has by then totally 
melted. 

The conclusion of this section, however, is that since super-cooled liquids behave at high frequencies 
($\omega \gg \gamma, \tau_\alpha^{-1}$)
like solids, the four-point correlation and dynamical susceptibility are expected to reveal, in a certain 
time domain, a non trivial behaviour unrelated to the structure of the `collective processes' discussed below 
(MCT, diffusive defects, CRR's) that one usually envisions to explain glassy dynamics.

\section{Mode Coupling Theory}\label {MCT}

As mentioned in the introduction the mode-coupling theory of supercooled
liquids predicts the growth of a cooperative length 
as the temperature is decreased or the density 
increased \cite{KT,FP,BB1}, and makes detailed predictions on the shape 
of $\chi_4(t)$. The four-point correlation function becomes critical near the mode-coupling transition temperature
$T_c$. The behaviour of the susceptibility $\chi_4(t)$ is encoded in ladder diagrams \cite{KT,BB1}. 
From the analytical and numerical results of \cite{BB1}, and analyzing 
the ladder diagrams \cite{BB1,BBB}, we have found that in the $\beta$ regime:
\be 
\chi_4(t)\sim f_1(t\epsilon ^{1/2a})/\sqrt{\epsilon}\qquad t \sim 
\tau_{\beta}
\ee
and in the $\alpha$ regime
\be
\chi_4(t)\sim f_2(t\epsilon ^{1/2a+1/2b})/\epsilon \qquad t \sim
\tau_{\alpha}
\ee
where $\epsilon=T-T_c$,
$a$, $b$ and $\gamma=1/2a+1/2b$ are the MCT exponents for the dynamical 
structure factor, and $f_1(x)$ and $f_2(x)$ 
are two scaling functions. Requiring that the dependence on $\epsilon$ drops out when 
$t\epsilon ^{1/2a} \ll 1$ one finds that 
$f_1(x)\sim x^a$ when $x \ll 1$. This leads to a power-law behaviour, 
$\chi_4 \sim t^a$, in the early $\beta$ regime, i.e. when the intermediate scattering 
functions approaches a plateau. In the same way, matching the behaviour of $f_1$ when $x \gg 1$ to the 
one of $f_2$ when $x \ll 1$ one finds another power-law behaviour, $\chi_4 \sim t^b$, 
on timescales between the departure from the plateau and the peak of $\chi_4$. We give in 
Fig.~\ref{mctfig} a schematic summary of the shape of $\chi_4(t)$ within the MCT description of
supercooled liquids.

Finally, as discussed in \cite{BB1}, at times $t =t^*  \sim \tau_\alpha$, $\chi_4$ 
reaches a maximum of height $(T-T_c)^{-1}$. Using the relation $\tau_\alpha \sim (T-T_c)^{-\gamma}$, 
valid within MCT, one finally finds $\chi_4(t^*) \sim t^{*1/\gamma}$. 

\section{Collectively Rearranging Regions} 
\label{CRR}
Under the term CRR, we gather many similar scenarii that differ in their 
details, as discussed in  the introduction \cite{AG,KTW,BB2,Tarjus}. Within the frustration-limited 
domains scenario of Ref.~\cite{Tarjus} it seems natural
to envision the dynamics as the activated motion of 
domains pinned by self-generated disorder. In the case of the random 
first-order theory of Refs.~\cite{KTW,BB2}, the details of the decorrelation mechanism 
are not entirely clear. There should be, on the one hand, activated fluctuations of domain walls 
between different states, again pinned by self-generated disorder. However, the fluctuations leading 
to a change of state may be the nucleation of a completely different state starting from the bulk. 
The latter process can be modeled as a nearly instantaneous event with a certain (small) nucleation rate. 
In the following we shall analyze separately these two types of fluctuations and their consequences on the
shape of $\chi_4(t)$.

\subsection{Instantaneous events}

Suppose that the dynamics is made of nearly instantaneous events that 
decorrelate the system in a compact `blob' of radius
$\xi_0$. The probability per unit time and volume 
for such an event to appear around site $\vec r$ is $\Gamma$.
We compute the four-body correlation of the persistence, $n_r(t)$, 
defined to be equal to one if no event 
happened at $\vec r$ between times 0 and $t$, and equal to zero 
otherwise. The four-body correlation is then defined as
\be \label{G4def}
G_4(\vec r,t)=\langle n_r(t) n_0(t) \rangle - \langle n_r(t) \rangle^2.
\ee
Clearly, the 
averaged correlation function, $C(t)=\langle n_r(t) \rangle$, is simply 
given by $C(t)=\exp(-\Omega \Gamma
\xi_0^d t)$ where $\Omega$ is the volume of the unit sphere.
For $G_4(\vec r,t)$ 
to be non zero, an event must have
happened simultaneously at $\vec r$ and at $0$, leading to
\be
G_4(\vec r,t)= C^2(t) \left[\exp\left(\Gamma t \xi_0^d f(r/\xi_0)\right) - 
1\right],
\ee
where $f(x)$ is the volume of the intersection between two spheres of unit 
radius with centers at distance $x$ apart.
Clearly, $f(x>2)=0$. Therefore, $G_4(\vec r,t)$ is non zero only if $r <2 
\xi_0$, and is in fact roughly 
constant there. For a given $r$ satisfying this bound, $G_4$ first grows 
linearly with time, reaches a maximum 
for $t = t^* \approx \Gamma^{-1} \xi_0^{-d}$ and decays 
exponentially beyond that time. The same behaviour is found for $\chi_4(t)$, 
that grows initially as $t^\mu$ with
$\mu=1$, and reaches a maximum such that $\chi_4(t^*) \propto \xi_0^d$. 
Assuming finally 
that these events are activated \cite{KTW,BB2}, with a barrier 
growing like $\Upsilon \xi_0^\psi$, 
where $\psi$ is a certain exponent, one expects $t^* \sim \tau_0 
\exp(\Upsilon \xi_0^\psi/T)$, and therefore 
$\chi_4(t^*) \propto (\ln t^*)^{d/\psi} \propto \xi_0^d$.

The rearranging regions could have of course more complicated shapes than 
the simple sphere assumed above. As
long as these objects are reasonably compact, the above results will still 
hold qualitatively. On the other hand,
if these regions are fractal with a dimension $d_f< d/2$, the above results 
on $G_4$ will hold with the argument in
the exponential given by $\Gamma t r^{2d_f-d}$; one also finds $t^* 
\approx 1/\Gamma \xi_0^{d_f}$ and 
$\chi_4(t^*) \propto \xi_0^{d_f}$.

\subsection{Domain wall fluctuations}

In this case the picture that we have in mind is similar to the case
of a disordered ferromagnet with pinned domain walls, where the typical 
time to flip a domain is comparable to
the inter-event time. In that case, an `event' is in fact the slow 
fluctuation of domain walls that progressively
invade the bulk of the domain. The early time behaviour of $\chi_4(t)$ is given by the square of 
the number of particles that relax per unit volume thanks to 
the same domain wall (see \cite{mayer} for the same situation out of
equilibrium in pure systems). Let again $\xi_0$ be the typical size of a domain
and  $\ell(t)$ the lengthscale over which 
the domain walls 
fluctuate during time
$t$. Considering that on the surface of each domain
there are order
$({\xi_0}/{\ell})^{d-1}$  subdomains of linear size
$\ell$  and that the number of particles in each of these subdomains
is proportional to $\ell^d$, we get $\chi_4(t) \propto \xi_0^{-d}
({\xi_0}/{\ell})^{d-1} \ell^{2d}\propto \ell^{d+1}/\xi_0$. We are descarding
for simplicity both the possibility of fractal domains and that transverse
fluctuations behave differently from longitudinal ones.
%$\xi_0^{d-2} \ell^2 \propto \xi_0^{d-2} (\ln t)^{2/\psi}$
Assuming thermal activation over  
pinning energy barriers that grow like $\Upsilon \ell^\psi$~\cite{FH}, we
finally get $\chi_4(t)\propto \xi_0^{-1} (\ln t)^{d+1/\psi}$. 
Therefore, in this case, the
exponent $\mu$ is formally zero and the growth of $\chi_4(t)$ is only 
logarithmic. The maximum of $\chi_4$ occurs
at time $t^*$ such that $\ell(t^*) \approx \xi_0$, 
which implies that the maximum of the susceptibility also
scales logarithmically with $t^*$, $\chi_4(t^*) \propto \xi_0^{-1} (\ln t^*)^{d+1/\psi} \propto \xi_0^d$.
The same scaling of the maximum of the susceptibility with the typical domain size is obtained in non-disordered 
coarsening systems~\cite{mayer}.

The conclusion of the above analysis is that if the CRR relaxation is due to both instantaneous events 
and domain wall fluctuations, the latter will dominate the time behaviour of $\chi_4$ before the peak as can be
readily deduced by comparing their relative contributions to $\chi_4(t)$. If for some reason, domain walls 
are particularly strongly pinned and bulk nucleation becomes dominant, then the exponent $\mu=1$ should be
observable. The height of the peak, on the other hand, behaves identically in both models. 
Thus, as the temperature is reduced, one should see a power-law behaviour before the peak with an 
exponent $0<b<1$ in the MCT regime followed by an effective exponent $\mu$ either decreasing towards
zero or increasing towards one depending on whether the domain wall contribution dominates or not. However, at
lower temperatures, the elastic contribution will also start playing a role, that might completely dominate
over the CRR contribution. This suggests that other observables, that quantify more specifically the collective
dynamics, should be devised to reveal a CRR dynamics.

\section{Defect mediated mobility}\label{defect}

\subsection{Independently diffusing defects}

As the simplest realisation of the defect mediated scenario for glassy
dynamics advocated in \cite{Jaeckle,FA,KA1,SR,GC}, 
we consider a lattice model in which mobility defects, or vacancies,
perform independent symmetric random walks. We assume for the moment that 
these vacancies cannot be created or destroyed
spontaneously. We shall compute the same function $G_4(\vec r,t)$ as in 
Eq.~(\ref{G4def}) above, arguing that when 
such a vacancy crosses site $\vec r$, the local configuration is
reshuffled and the local correlation drops to zero. Therefore,
$n_r(t)$ is equal to one if no vacancy ever visited site  $\vec r$ 
between $t=0$ and $t$, 
and zero otherwise. Thus, $\langle n_r(t) \rangle $ represents a
density-density dynamical correlation function whereas 
$\langle  n_0(t)n_{\vec{r}} (t)  \rangle -\langle  n_0(t)\rangle^{2}$ 
corresponds to $G_4(\vec r,t)$.  

From now on we will denote by $N_v$ the number of vacancies, by $V$  the
total volume, by $\rho_v=N_v/V=1-\rho$ the vacancy density  and by 
$P^z_{{\overline x}}(t)$ the
 probability that a vacancy starts in $z$ at time zero
and never reaches $x$ till time $t$. 
The probability that a vacancy starts in $z$ at time zero
and reaches for the first time $x$ at a time $u\leq t$
is therefore  $P^z_{x}(t)=1-P^z_{{\overline x}}(t)$.

The computation of $\langle n_x(t) \rangle$ is identical to the target
annihilation problem considered in \cite{Redner}.
Since we assume 
defects to be independent, the defect distribution is uniform and we have:
\be
\label{1}
\langle n_x(t) \rangle = \left[ \frac{1}{V}
\sum_{z,z\neq x} P^z_{{\overline x}}(t) \right]^{N_v}=
\left[ \frac{1}{V} \sum_{z,z\neq x} (1-P^z_{x}(t)) \right]^{N_v}=
\exp \left[ -\rho_v-\rho_v \sum_{z,z\neq x} P^z_{x}(t) \right].
\ee

The correlation function $\langle n_x(t)n_y(t) \rangle$ can be also
expressed in terms of probability distributions of a single random walk in a
similar way: 
\begin{eqnarray}
\label{2}
\langle n_x(t)n_y(t) \rangle &=&
\left[ \frac{1}{V}\sum_{z,z\neq x,y} P^z_{{\overline x}, {\overline y}}(t)
\right]^{N_v}=
\left[ \frac{1}{V}
\sum_{z,z\neq x,y} (1-P^z_{x}(t)-P^z_{y, {\overline x}}(t)
\right]^{N_v}=\nonumber\\
&&\left[1-\frac{2}{V}-\frac{1}{V}
\sum_{z,z\neq x,y} P^z_{x}(t)-\frac{1}{2V} 
\sum_{z,z\neq x,y}(P^z_{y, {\overline x}}(t)+P^z_{x, {\overline y}}(t))
\right]^{N_v}=\nonumber\\
&&\exp \left(-2\rho_v-\rho_v\sum_{z,z\neq x} P^z_{x}(t)+\rho_v
  P^y_x(t)-\frac{\rho_v}{2}\sum_{z,z\neq x,y} (P^z_{y,{\overline x}}(t)+
P^z_{x,{\overline y}}(t))\right),
\end{eqnarray}
where $P^z_{{\overline x}, {\overline y}}(t)$ is the 
probability that a vacancy starts in $z$ at time
zero and never reaches neither $x$ nor $y$ till time $t$,
$P^z_{x, {\overline y}}(t)$ is the probability 
that a vacancy starts in $z$ at time zero and reaches $x$ at $u \leq t$ 
but never reaches $y$ till time $t$.
In Eqs. (\ref{1},\ref{2}) we are left with 
the calculation of probabilities of the form
$P^z_x(t)$, $P^z_{x, {\overline y}}(t)+P^z_{y {\overline x}}(t)$ for
a single random walk. This can be done using Laplace transforms
and, concerning $P^z_x(t)$, the computation has been performed
a while ago \cite{Montroll}. All the details can be found in Appendix B.

In the continuum limit, $(x-y)/\sqrt{Dt/2}\sim O(1)$, 
i.e. for independent Brownian motion with diffusion
coefficient $D$, the final expression for $\langle n_{x} (t)\rangle$ on time 
scales much larger than one is, in three dimensions,
\be
\langle n_{x} (t)\rangle=\exp [-\rho_{v}-c_{1} D \rho_{v}t],
\ee
where $c_{1}$ is a constant fixed by the short-lengthscale physics,
i.e. the underlying lattice structure (see Appendix B). It is clear 
from this expression which is valid in all dimension larger than two 
that the relaxation time-scale 
is governed by the vacancy density $\rho_{v}$ and reads 
$\tau =1/(c_1 \rho_{v}D)$. Physically $\tau$ 
corresponds to the time such that each site has 
typically been visited once by a defect. 

The final expression for $G_{4}$ is, for time and length scales much
larger than one, and in the small vacancy density limit, 
$\rho_{v}\rightarrow 0$,
\begin{equation}\label{g43db}
G_{4}(\vec{r},t)=\frac{c_{2}}{\rho_{v}}\exp \left(-\frac{2t}{\tau }\right)
\left(\frac{t}{\tau } \right)^{2}\int_{0}^{1} du \int_{0}^{u}dv
\frac{e^{-\frac{r^{2}}{2Dvt }}}{(2Dvt )^{3/2}},
\end{equation}
where $c_{2}$ is a constant of order unity.
Note that the correlation length at fixed $t$ is given by $\xi(t)=\sqrt{Dt}$. 
For $r \ll \xi(t)$, $G_4(\vec r,t)\sim 1/r$ whereas for $r \gg \xi(t)$, 
$G_4$ decays at leading order as a Gaussian, that is, much
faster than exponentially. 
The $1/r$ behaviour is cut-off on short-length scales, where (\ref{g43db})
does not hold. For $r=0$ one finds, when $t \gg 1$,
\be
G_{4}(r=0,t) = \langle n_x(t) \rangle - \langle n_x(t) 
\rangle^2 = \exp(-t/\tau) \left[1 - \exp(-t/\tau) \right], 
\ee
which behaves as $t/\tau$ at small times.

By integrating (\ref{g43db}) over $\vec r$ we get the dynamical susceptibility,
\be
\chi_{4} (t)=\frac{c_{2}}{2\rho_{v}} 
\left(\frac{t}{\tau}\right)^{2}\exp\left(-\frac{2t}{\tau }\right).
\ee
For short times, $t<\tau $, the dynamical susceptibility is
proportional to $t^{2}$, so that 
$\mu=2$. This is due to the diffusing nature of the
defects. The main contribution to $\chi_{4}$ is given by the square
of the number of sites visited by the same defect, which behaves as
$\rho_{v} (Dt)^2=\frac{1}{\rho_{v}}\left(\frac{t}{\tau}\right)^{2}$, 
since a random walk in 
three dimensions typically visits $t$ different sites. 
For $t > \tau$, on the other
hand, the correlation decreases because sites start being visited by different
vacancies. 
The maximum of $\chi_4(t)$ is reached for $t=t^*=\tau$, for which one has
$\chi_4(t^*) \sim \rho_{v}^{-1} \sim Dt^*$. 
%As a consequence if $D$
%depends on $\rho_v$, as it happens for example for the one-spin
%facilitated FA model where $D\propto 1/\sqrt{t^{*}}$, then the power law
%dependence of the maximum of $\chi_4$ is different from $t^{*}$. 
Note that because random walks are fractals of dimension 
$d_f=2$, 
the above relation can also be written as $\chi_4(t^*) \sim a^{d-d_f} \xi^{d_f}(t^*)$, 
where we have added the lattice spacing $a$ to give to $\chi_4$ the dimension of a volume.
If for some reason $D$ depends on $\rho_v$,  
as it happens for example for the one-spin facilitated FA model where $D \propto \rho_v$, 
then one finds $t^* \sim \rho_v^{-2}$ and $\chi_4(t^*) \sim t^{*1/2}$.

Taking the Fourier transform of $G_{4}(r,t)$ given by Eq. (\ref{g43db}), 
we find the four point structure factor,
$S_4(k,t)$,
\be\label{s4k3d}
S_4(k,t) = \chi_4(t) {\cal F}(Dk^2 t); \qquad {\cal F}(u) 
\equiv \frac{2}{u^2} \left(u-1+e^{-u}\right).
\ee
Note that $S(k=0,t)=\chi_4(t)$, as it should. Furthermore for large
and small $k$, $S_4(k,t)$ behaves respectively as $S_4 \sim k^{-2}$ and 
$S_4 \sim \chi_4 + O (k^{2})$, just as the 
Ornstein-Zernike form, though the detailed $k$ dependence is different. 

One can also study this problem in dimensions $d=1$ or $d=2$. 
Qualitatively, the same conclusions
hold (diffusive correlation length $\sqrt{Dt}$, correlation time 
$t^*$ set by the density 
of vacancies, etc.), although the quantitative results differ because 
a random walk in $d \le 2$ 
visits a number of sites that grows sub-linearly with time, see
Appendix B.1 and B.3. One finds in particular that
$\chi_4(t^*) \sim (Dt^{*})^{d/2} \sim \xi^{d}(t^*)$, with logarithmic 
corrections for $d=2$. 
The above arguments can be generalized if for some reason the 
vacancies have an 
anomalous diffusion motion, in the sense that their typical excursion 
between time $t=0$ and 
time $t$ scales as $t^{1/z}$, where $z$ is the dynamical exponent. When 
$z = 2$, usual diffusion 
is observed, but many models like diffusion in random media or kinetically 
constrained models may lead to sub-diffusion, where 
$z > 2$ \cite{WBG,Bertin}. In this case, one expects the small time behaviour of 
$\chi_{4} (t)$ to be given by
$\chi_{4} (t) \sim t^{2d/z}$ for $d < z$ and $t^2$ for $d > z$ with 
logarithmic corrections for $d=z$.
Similarly, the behaviour of $\chi_4(t^*)$ is a power-law, $\chi_4(t^*) 
\sim t^{*\lambda}$, with 
$\lambda=d/z$ for $d < z$ and $\lambda=1$ for $d > z$.

In the above model, mobility defects were assumed to be conserved in time. However, it is certainly more 
realistic to think that these defects can be spontaneously created and disappear with time. Suppose that
defects are created with a rate $\Gamma$ per unit time and unit volume, and disappear with a rate $\gamma$
per unit time. The equilibrium density of defects is then $\rho_v=\Gamma/\gamma$. The above results on 
$\chi_4$ can easily be generalized. At small times, the number of pairs of visited sites will now behave
as $\rho_v (Dt)^2 -\frac 2 3 \Gamma (Dt)^3/D$. Because of the death of vacancies there is an extra decay of the
dynamical susceptibility. The dominant rate of decay depends on the adimensional number $\gamma \tau$.

A very similar model for glassy dynamics was suggested in \cite{Houches}, where free volume is described as a diffusing 
coarse grained density field $\rho(\vec r,t)$ with a random Langevin noise term. Mobility of particles is allowed whenever
the density $\rho$ exceeds a certain threshold $\rho_0$. The mobile regions are then delimited by the
contour lines of a random field, which already gives rise to a quite complex problem of statistical geometry 
\cite{Isichenko}. The particle density correlation in this model is a 
simple exponential with relaxation time $\tau \sim \exp(\rho_0/{\overline \rho})$, where $\overline{\rho}$ is the 
average free volume density. One can also compute $\chi_4(t)$ in this model to find, in $d=3$, 
\be
\chi_4(t) \sim t \left[\exp \left(-\frac{t}{\tau} \right) 
\left(1-\exp\left(-\frac{t}{\tau}\right) \right) \right]
\ee
which behaves very much like the point like vacancy model studied above, with
in particular, $\chi_4(t) \sim t^2$ for $t \ll \tau$. 

Let us finally note that from the point of view of interacting particles on a lattice
we have studied the persistence dynamical susceptibility, instead of the 
density-density correlations discussed in the introduction. This is because 
for the lattice 
gas problem at hand, the former does not show any interesting
properties: except when a defect passes by, the local state is always
the same, i.e. occupied. For completeness, we give the corresponding 
results in Appendix B.4.   In a real system, however, 
the local configuration is going to be
affected by the passage of a mobility defect, and one can expect that the 
density-density correlations 
will in fact behave more like the persistence dynamical susceptibility
computed before.
The correspondence between persistence and self-intermediate
scattering function is studied explicitly in 
kinetically constrained models in Ref.~\cite{BCG}.

\subsection{Kinetically Constrained Models: numerical results}

Kinetically constrained models (KCM) postulate that glassy dynamics
can be modeled by forgetting about static interactions
between particles, putting all the emphasis on dynamical aspects.
Among those models are, for example, the Fredrickson-Andersen (FA), or
the Kob-Andersen (KA) on hyper-cubic lattices \cite{SR,TBF}.
% with one vacancy or 
%more needed to facilitate the motion 
%of other vacancies . 
The dynamics of these models can be
understood in terms of diffusion of defects \cite{Sollich,WBG,TBF}
and the models can be classified into cooperative and non-cooperative models,
depending on the properties of such defects. 
For cooperative models 
%with more 
%constrained dynamics 
the size, $\xi_0$, the density, 
and the time-scale for motion of the defects depends on the particle
density (for conservative models) or temperature (for non-conservative models)
and change very rapidly with increasing density or decreasing temperature 
~\cite{TBF}. KA and FA models with more than one
neighbouring vacancy
needed in order to allow the motion of other vacancies belong to this class. 
On the other hand
for the one-spin isotropically facilitated 
FA model, a single facilitating spin is a mobile defect at all 
values of temperature and the model is non-cooperative. A recent analysis \cite{WBG} suggests 
that for these models defects can be considered as non-interacting 
in $d>4$, while for $d < 4$ the r\^ole of fluctuations becomes important.
Therefore we expect that the previous results for the independent diffusing
defects model should
apply exactly for FA one-spin facilitated in $d>4$.
Furthermore, since
the corrections to the Gaussian exponents are not very large \cite{WBG} 
in three dimensions, we still expect a semi-quantitative agreement.
In particular the initial 
increase of the dynamic susceptibility as 
$\chi_4(t) \sim N(t)^{2}$, where $N(t)$ is the total number 
of distinct visited sites, is expected to be quite a robust result.
Also, we expect a diffusive growth of the dynamical length 
scale $\xi(t)$ governing the scaling of $G_4$, at least in the 
limit $\xi(t)\gg \xi_0$. At smaller times, one expects a 
crossover between a CRR regime when $Dt \ll \xi_0^2$ (where the
dynamics inside the defects becomes relevant in cooperative models
to a mobility defect regime 
for longer times. 
Hence, in principle, looking at the detailed properties of $G_{4}(r,t)$ 
one should be able to extract the defect properties: density, size, 
time-scale and decide which 
theoretical scenario is most consistent with numerical results. 

In the following, we discuss numerical results for
the one-spin facilitated FA  model both in 
$d=1$ and $d=3$, and for the $d=1$ East model where 
facilitation is anisotropic \cite{SR}.  The two models can be
described respectively in terms of diffusive and sub-diffusive
non-cooperative defects and indeed the numerical results are in
quantitative agreement with the predictions of previous section, as will be
explained in detail. We do not address the case of cooperative KCM models,
for which a more complicated behaviour is
expected. Indeed a first slowing down of dynamics should occur
near a dynamical crossover displaying the properties of an 
MCT-like avoided transition \cite{TBF}. 
In this regime the model cannot be approximated 
as  a system of independent freely diffusing defects and 
deriving a quantitative prediction 
for the behaviour of four point correlation
and susceptibility would deserve further work.
Such avoided transition should then be
followed at lower temperature or higher density
by an asymptotic behaviour described in terms of cooperative diffusing defects.

\subsubsection{One dimension}

Let us start with the simplest model, the $d=1$ FA model. 
For a given temperature, we consider 
the time evolution of the following quantities. The analog of the
spatial four-point correlator for this model is
\be
G_4(r,t) = \frac{1}{N} \sum_{i=1}^N \left[
\langle n_i(t) n_{i+r}(t) \rangle - n^2(t) \right],
\label{c4}
\ee 
where
$n(t) = N^{-1} \sum_{i=1}^N \langle n_i(t) \rangle$
is the mean persistence, $n_i(t)$ being the persistence 
at site $i$.
We also measure the corresponding four-point structure factor
\be
S_4(k,t) = \frac{1}{N} \sum_{\ell,m=1}^N \left[ \langle n_\ell(t) n_m(t)
\rangle  -n^2(t) \right]
e^{i k \cdot (\ell-m)},
\label{s4}
\ee
and as usual we get the four-point susceptibility as the 
$k \to 0$ limit of the structure factor, $\chi_4(t) = S_4(k=0,t)$.
We generally find that the results are in good agreement with the free
defect model described above, at least at sufficiently 
low temperatures.  

\begin{figure}
\psfig{file=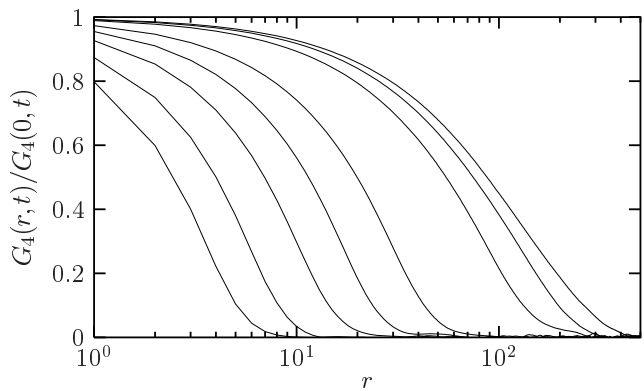,width=8.5cm}
\caption{\label{1dfa} 
Four-point spatial correlator (\ref{c4}) in the $d=1$ FA model
at fixed temperature, $T=0.2$, and various 
times, $t= 10^3$, $3.10^3$, 
$10^4$, $3.10^4$, $10^5$, $10^6$, $3.10^6$, $6.10^6$
(from left to right). 
The correlator is normalized by its $r=0$ value.
At this temperature, the relaxation time is 
$\tau \sim 10^6$, so that time-scales cover both regimes where
$t/\tau$ is smaller and larger than 1.}
\end{figure}

In Fig.~\ref{1dfa}, we show the evolution of the 
spatial correlator (\ref{c4}) at a given low temperature, 
$T=0.2$, and various times. At this temperature, 
the relaxation time is about $\tau \sim 10^6$, so that 
time-scales presented in Fig.~\ref{1dfa} cover a range 
of times both smaller and larger than $\tau$. 
The dynamic susceptibility $\chi_4(t)$ has the usual shape with 
a maximum at a time close to $\tau$ indicating that 
dynamics is maximally heterogeneous there. 
This non-monotonic behaviour of $\chi_4$ in fact does not show up 
in the spatial correlators of Fig.~\ref{1dfa}, which display
instead a smooth monotonic evolution with time. The spatial decay 
of $G_4(r,t)$ becomes slower when $t$ increases indicating the presence
of a monotonically growing dynamic lengthscale $\xi(t)$.  

One can estimate the time dependence of $\xi(t)$ by 
collapsing the data of Fig.~\ref{1dfa} using a form like:
\be
G_4(r,t) \sim G_4(0,t) \, {\cal G} \left( \frac{r}{\xi} \right).
\label{scal1d}
\ee
Doing so, we find 
that $\xi \sim t^{0.45}$ is a reasonable representation of the data
at $T=0.2$. Correspondingly, we find that the increase of 
$\chi_4(t)$ for $t < \tau$ is well-described by a power-law, 
$\chi_4 \sim t^{0.85}$, so that the expected 
scaling $\chi_4 \sim \xi^2$ is reasonably verified
given the unavoidable freedom in estimating the range of time-scales 
where power-laws apply. 
The values of these exponents 
are not far from the ones expected from freely diffusing defects
in one dimension, although slightly smaller. Indeed, we recall that the
results in Appendix B.1. predict $\xi=\sqrt{Dt}$, $\chi_4(t)\propto
\rho \xi(t)^2$ and $\chi_4(t^{*})=1/\rho$, where $\rho$ is the density of
defects, $D$ their diffusion coefficient and $t^{*}$ the time at which
$\chi_4(t)$ reaches its maximum value. This last prediction is also in good
agreement with the numerical results (see e.g. \cite{GC}).

Repeating the simulation at lower temperature, $T=0.15$, we obtain 
$\chi_4 \sim t^{0.93}$, showing that deviations from theoretically 
expected values are partly due to preasymptotic effects
that presumably disappear at very low temperatures.

It is important to remark that the scaling form 
(\ref{scal1d}) is only approximately supported by the data.
The scaling 
in fact deteriorates when times become larger than $\tau$. 
This can be seen in Fig.~\ref{1dfa} where
data for large times become more and more stretched, indicating an
increasing polydispersity of the dynamical clusters. 
Note that a change in 
the shape of the spatial correlator makes a quantitative determination
of $\xi$ problematic. Usually, one wants to collapse various
curves using a form like (\ref{scal1d}) to numerically 
extract $\xi$. Strictly speaking, this is not possible 
here if one works at fixed $T$ and varying $t$
over a large time window. 
This difficulty provides a second possible explanation for the small
discrepancy between the measured values of exponents 
and the theoretical expectations.

The observation of a monotonically growing length begs the 
question: how can
the correlation length increase monotonically with time
while the volume integral of the spatial correlator $\chi_4$
is non-monotonic, as reported in the previous section? This is due to the fact
that we have presented in Fig.~\ref{1dfa} results for the normalized 
correlator, $G_4(r,t)/G_4(r=0,t)$. By definition, $G_4(0,t) = n (1-n)$, hence
the normalization itself exhibits a non-monotonic behaviour. 
If one considers the normalized susceptibility, 
$\tilde{\chi_4} = [G_4(0,t)N]^{-1}  \sum_{\ell,m} \left[ 
\langle n_\ell n_m \rangle - n^2(t) \right]$, 
one indeed finds that $\tilde{\chi_4}$ is monotonically growing as well.

In numerical works, the quantities that have been studied are in fact, 
most of the time, normalized, 
and the corresponding $\tilde{\chi_4}(t)$ observed for realistic systems 
shows a peak, at variance
with what is observed in the $d=1$ FA model. As we shall show below, 
this is due to the one-dimensional
nature of the model, and this difference is not observed 
in three dimensions. This difference 
in the behaviour of the normalized dynamical susceptibility 
between one and three dimensions is indeed in full agreement
with the independent defect diffusion computation, see previous 
section and Appendix B. 

Results are qualitatively similar in the one-dimensional East model. 
The dynamic susceptibility $\chi_4(t)$ 
develops a peak that grows and whose position
is slaved to the increasing relaxation time
when temperature decreases. 
At fixed temperature, a monotonically 
growing lengthscale is observed, while the scaling relation
$\chi_4 \sim \xi^2$ still holds within our numerical precision.
The novelty of this model lies in the fact that exponents are now 
temperature dependent, 
as all other dynamic exponents in this model.
%Moreover, the values extracted from the numerical simulations
%are far from the freely diffusing defects predictions. 
For instance, we find that $\xi(t) \sim t^{0.28}$ at $T=0.4$,  
$\xi(t) \sim t^{0.15}$ at $T=0.2$. 
These results are in agreement with the above predictions
of the independent defects
model if the defect motion is
%in the East model can be  
%interpreted as being 
sub-diffusive, with a dynamic exponent $z=T_0/T$, 
as expected from \cite{Sollich}. 
Due to the quasi one-dimensional nature of the relaxation process  
in the three-dimensional generalization of the East model~\cite{nef},  
these results most probably carry over to larger dimensions 
where they would differ by numerical factors only.   
 
\subsubsection{Three dimensions} 
 
In $d=3$, the situation is more subtle. 
Results for the normalized susceptibility of the one-spin facilitated FA model 
were presented in Ref.~\cite{WBG2}, where it was found 
to have the standard non-monotonic shape already described several times above. 
We find that the non-normalized $\chi_4(t)$ has the same qualitative behaviour.   
Therefore, contrary to the $d=1$ case normalization is not a crucial issue in three dimensions. 

In the following we check the predictions for independent diffusing defects in
three dimensions for the susceptibility and correlation length obtained
above, i.e. $\xi(t)=\sqrt{Dt}$, $\chi_4(t)\propto
\rho~\xi(t)^4$ and $\chi_4(t^{*})=1/\rho$,  where $\rho$ is the density of
defects, $D$ their diffusion coefficient and $t^{*}$ the time at which
$\chi_4(t)$ reaches its maximum value. We find a semiquantitative agreement
with above prediction, with small deviations in the exponents that should be due to the interaction among
defects. In particular the scaling of the peak with the
density of defects was already analyzed in \cite{WBG2}, where the result
$\chi_4(t^{*})\propto 1/\rho^{1-\epsilon}$ was obtained, with
$\epsilon\simeq 0.03$.
As for the correlation length, we find $\xi(t)\propto t^{0.42}$, that shows 
again a small deviation from
the diffusive prediction. Regarding the increase at
$t\ll\tau$ of the susceptibility we find a power-law as predicted.
As in $d=1$, the exponent changes slightly when 
decreasing temperature because the scaling regime where 
power-law applies becomes more and more extended. 
We find $\chi_4 \sim t^{1.4}$ at $T=0.25$, 
$\chi_4 \sim t^{1.55}$ at $T=0.17$ and $\chi_4 \sim t^{1.89}$ at 
$T=0.095$. 
This seems to indicate that the deviation from the 
scaling $\chi_4(t)\propto t^2$ calculated for the independent diffusing
defect model is partly due to preasymptotic effects
that are less and less important at lower temperature. 
Unfortunately, 
we were not able to measure $\xi$ at much lower temperatures
with sufficient accuracy. We expect that even at very low temperature a small
deviation from the exponent of independent defects should survive due to the interaction among defects.
\begin{figure}
\psfig{file=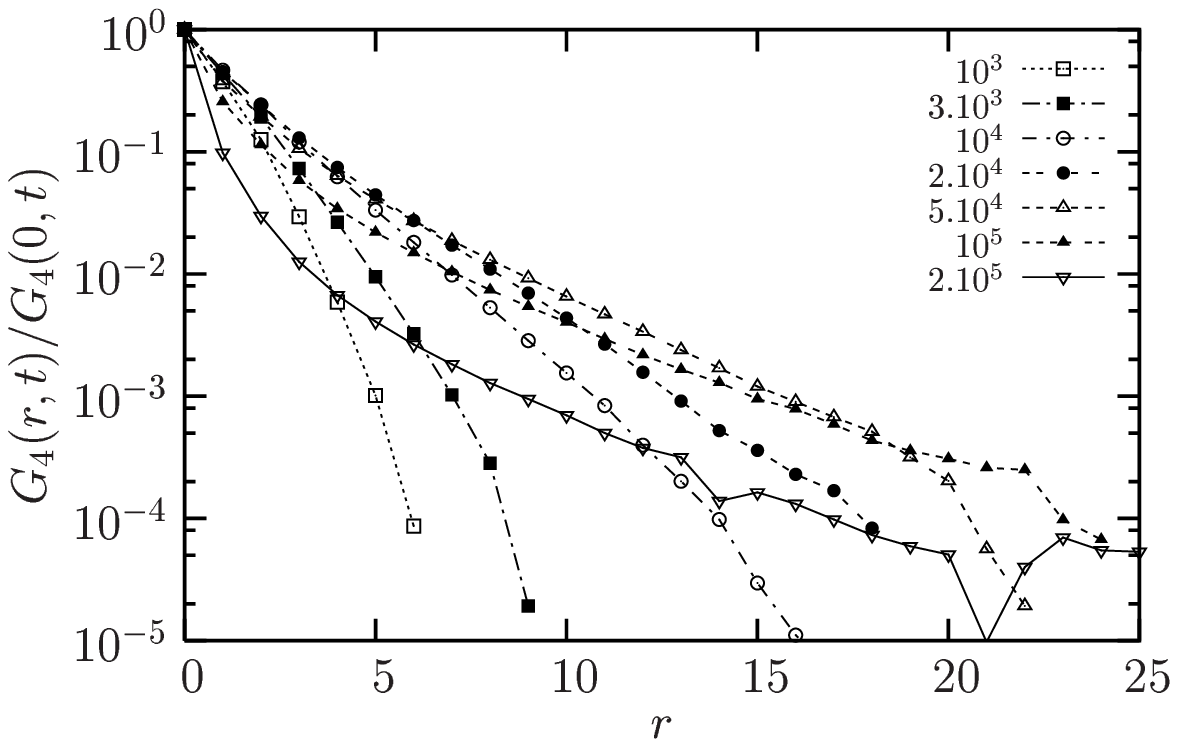,width=8.5cm}
\psfig{file=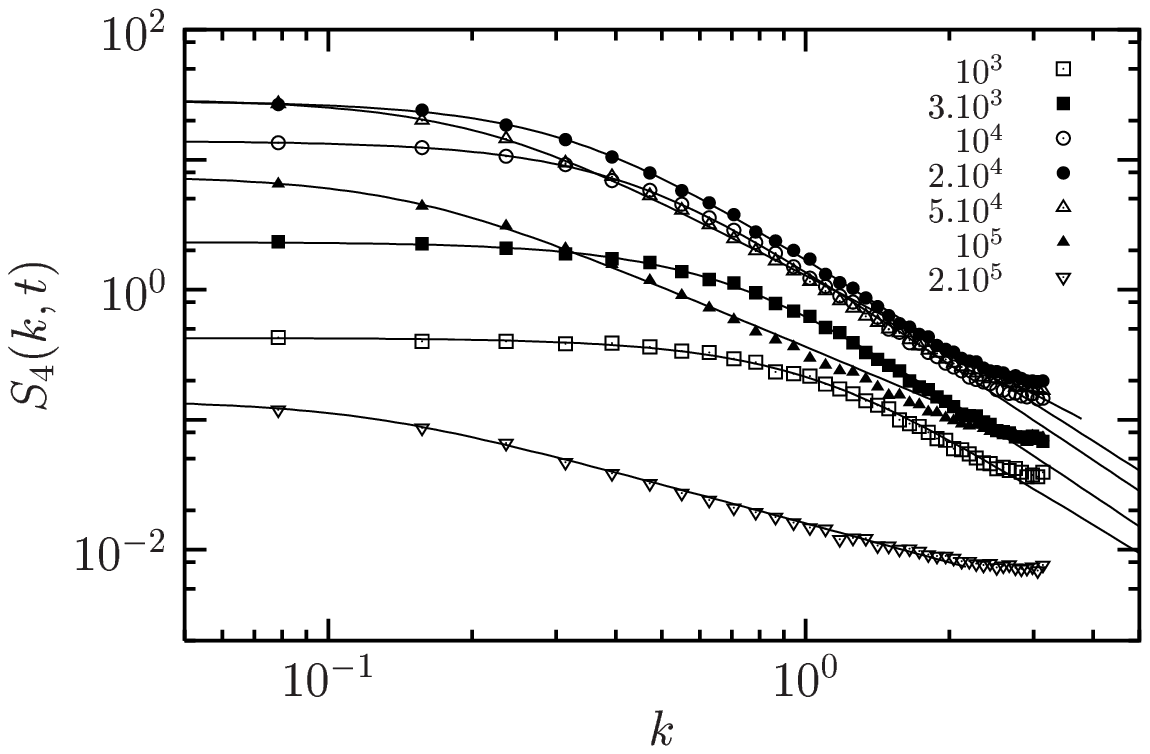,width=8.5cm}
\caption{\label{3dfa2}
Four-point correlations in the $d=3$ one-spin facilitated FA model
both in real space (left) and in Fourier space (right)
at fixed temperature, $T=0.17$, 
and various times indicated in the figures.
In Fourier space, points represent numerical data, 
while full lines are fits to the form (\ref{fit3d})
with fitting parameters described in the text.}
\end{figure}

In Fig.~\ref{3dfa2} we show the four-point 
correlations both in real and Fourier space, Eqs.~(\ref{c4}) and (\ref{s4}).
In these curves 
the temperature is fixed at a low value, $T=0.17$, and time is varied in a
wide range that includes the relaxation time, $\tau(T=0.17) \sim 5.10^4$, where 
the dynamic susceptibility also peaks.
For times $t \ll \tau$, the spatial decay of $G_4(r,t)$ is fast. When 
$t$ increases, the spatial decay becomes slower, 
once again indicative of an increasing dynamic correlation
length $\xi(t)$. When $t$ becomes larger than $\tau$, however,
spatial correlations seem to become weaker. 
It is obvious from Fig.~\ref{3dfa2} that the volume integral
of $G_4(r,t)/G_4(0,t)$ decreases when $t$ grows larger than $\tau$.
This is very different from the one-dimensional case in Fig.~\ref{1dfa}, 
but consistent with all known numerical results.  
 
However, a closer look at Fig.~\ref{3dfa2} reveals that 
even though the initial spatial decay of $G_4(r,t)$ is stronger at  
larger times, the contrary is true at large distances.  
This indicates that the topology of the dynamic clusters 
changes when $t$ grows larger than $\tau$, but that  
$\xi(t)$ may keep increasing in a monotonic manner. 
Since the spatial correlator is very small at large distances, 
quantitative measurements of $\xi(t)$ are more easily 
performed in Fourier space via $S_4(k,t)$. 

At short time, a fit of $S_4(k,t)$ 
using the functional form given by Eq. (\ref{s4k3d}) 
works reasonably well, but the fit quickly deteriorates 
at long time. We have therefore used the following generalization of 
Eq.~(\ref{s4k3d}):
\be
S_4(k,t) = \chi_4(t) {\cal F}_\beta[k^2 \xi^2(t)]; 
\qquad {\cal F}_\beta(u) \equiv \frac{2^{2/\beta}}{u^\beta} 
\left(u-1+e^{-u}\right)^{\beta/2}.
\label{fit3d}
\ee
Freely diffusing defects  correspond
to $\beta=2$ and $\xi(t) \sim \sqrt{t}$. Using $\beta(t)$ as an
additional free parameter, we are able to fit $S_4(k,t)$ at all times, see 
Fig.~\ref{3dfa2}.
We find that $\beta$ decreases from $\beta \approx 2.5$ 
at small times to $\beta \approx 1$ for the longest 
times scales investigated, which corresponds to 
$t \approx 5 \tau$. At 
such large times, the dynamic susceptibility 
has already decreased by a factor $\approx 300$ from its maximum 
value at $t = \tau$, and correlations become very weak indeed.
The values for $\beta$ found from the fits are consistent
with the value $\beta \approx 2.15$ reported in Ref.~\cite{WBG2}
where only fixed time ratio $t/\tau(T)=1$ 
at different temperatures have been studied.
From this fitting procedure, 
we deduce a monotonically growing dynamic length $\xi(t)$, even beyond $t =\tau(T)$. 
Fitting its time dependence with a power-law, we get
$\xi \sim t^{0.42}$ which appears to be slightly
sub-diffusive, but close to the value found above in the one-dimensional case.

In conclusion we find that on small
enough time-scales, one indeed has good 
agreement with the above calculations based
on freely diffusing defects,
therefore defect branching 
and defect coagulation can be neglected. 
However, for longer time-scales, significant deviations 
appear which correspond to the evolution of the exponent $\beta(t)$ and should
be responsible for the small deviations of the predicted exponent for
$\chi_4$.
Physically, the time evolution of the exponent $\beta(t)$ characterizing
the large $k$ behaviour of the dynamic structure factor is reasonable. 
At very short-times, dynamic clusters consist of coils 
created by random walkers, and an exponent close to 
$\beta=2$ can be expected. For times $t \sim \tau$, clusters look 
critical, as described in Refs.~\cite{WBG,WBG2}, and the exponent 
$\beta=2-\eta$, $\eta < 0$ is expected. 
At very large times, clusters are most probably extremely polydisperse
because the remaining spatial correlations at large times 
are due to the largest regions of space that were 
devoid of defects at time 0 and that take therefore a large time 
to relax. But at large times, some isolated sites 
that have not been visited by defects during the relaxation
might survive so that
the distribution of dynamic clusters at large times is very wide, 
see Ref.~\cite{nef} for snapshots.
A small value of $\beta$ can therefore be expected.

\section{Numerical results on atomistic model systems}
\label{numerics}
In this section, we study numerical results 
for the dynamic susceptibility and structure factor
of a super-cooled liquid simulated by molecular dynamics simulations. 
The model we study is mainly the well-known binary Lennard-Jones mixture 
as first defined and studied in Ref.~\cite{KA}, but we report also some
results for a soft-spheres mixture studied in \cite{Barrat,Grigera,Dave}. 
We do not give details about our 
numerical procedures since these 
were given several times in the literature~\cite{KA,Berthier2,WBG}.

\subsection{Dynamical susceptibility}

\begin{figure}
\psfig{file=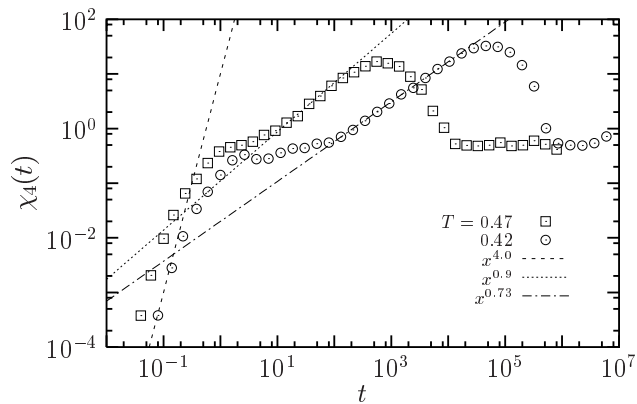,width=8.5cm}
\caption{\label{lj1} Time dependence of the dynamic susceptibility 
in the binary LJ mixture at two different temperatures. The lines
are power-law fits with exponents indicated in the label.}
\end{figure}

In previous works on various realistic liquids, the 
dynamic susceptibility was reported several 
times~\cite{Onuki,Glotzer,FP,parisi}. 
It is 
known to exhibit at peak at a time-scale slaved to 
the quantity chosen to quantify local dynamics. Typically, 
particle displacements are chosen, and one computes 
therefore the variance of some dynamical correlation, 
\be
\chi_4(t) = N \left[ \langle F^2(t) \rangle - \langle F(t) \rangle^2 \right],
\ee
with 
\be
F(t) = \frac{1}{N} \sum_{i=1}^N F_i(t). 
\ee
The dynamic quantity $F_i(t)$ can be chosen as some 
`persistence' function in which case $\langle F(t) \rangle$ resembles
the overlap function usually measured in spin 
systems~\cite{Glotzer,FP,parisi}. 
Other choices 
are~\cite{WBG,Berthier2}
\be
 F_i(t) = \cos( \vec{q} \cdot \delta \vec{r}_i(t)),
\label{fk}
\ee
where $\vec{q}$ is a wavector chosen in the first Brillouin zone, and 
$\delta \vec{r}_i(t)$ is the displacement of particle $i$ in a time 
interval $t$. In the limit of small $|\vec{k}|$, it is better to study
$F_i(t) = |\delta \vec{r}_i(t) | / \sqrt{\Delta r^2(t)}$, where 
$\Delta r^2(t)$ is the mean square displacement of the 
particles~\cite{Onuki,heuer}.

Whereas the general shape of $\chi_4(t)$ is well-documented in
the literature, its precise time dependence was never discussed. 
In Fig.~\ref{lj1}, we present the time dependence of $\chi_4(t)$
in the binary Lennard-Jones mixture at two different temperatures. 
The data are presented in a log-log scale, in order to emphasize
the existence of several time regimes that are generally hidden
in the existing reports. 
To build these curves, we choose (\ref{fk}) as the local 
observable, for a wavector that 
corresponds roughly to the typical inter-particle distance. 

In the ballistic regime at very short-times, we find that 
$\chi_4(t) \sim t^4$, as described from Section~\ref{Short-time behaviour}. 
The system then enters the time regime where 
dynamic structure factors typically exhibit plateaus, as 
a result of particle caging. As seen in Fig~\ref{lj1}, 
this is also the case for $\chi_4(t)$. 
Finally, $\chi_4(t)$ reaches a maximum located close to the 
relaxation time extracted from the time dependence of $\langle F(t) \rangle$, 
and then rapidly decays to its long-time limit, equal to $1/2$ in the present 
case. 
In Fig.~\ref{lj1}, we fitted the time dependence of 
the increase of $\chi_4(t)$ towards its maximum with power-laws $\chi_4 \sim t^\mu$. 
The fits are satisfactory, although they only hold on 
restricted time windows. We find a slight temperature dependence
of the exponent $\mu$. For instance, we find $\mu \approx 0.9$ at
$T=0.47$, and $\mu \approx 0.73$ at $T=0.42$.
As already discussed in the case of kinetically constrained models
above, it is not clear how the restricted time 
window used to determine the exponents might affect their values.
However, the data in the Lennard-Jones system
behave quantitatively very differently from both 
theoretical results obtained from 
freely diffusing defects and numerical results 
in the one-spin facilitated $d=3$ FA model, where $\mu=2$. 
The small temperature evolution in the LJ liquid 
differs even qualitatively from the one-spin facilitated $d=3$ FA model
where the exponent was found to increase when decreasing temperature. 
These observations tend to discard a description of this 
super-cooled liquid via a scenario with simple independently diffusing
defects, even interacting ones. The above value of $\mu$ is in principle 
compatible with the predictions of elasticity theory, which yields $\mu=1/2$ or
$\mu=1$ depending on the damping of phonons. However, the time scale in which the above
mentioned power-law behaviour holds in the Lennard-Jones mixture corresponds to the $\beta$
regime where the displacement of particles is no longer small and the elastic description
unjustified. Within MCT, on the other hand, 
$\chi_4$ should increase in that regime with an exponent $\mu=b$
that is known from previous analysis of the dynamics
of the binary Lennard-Jones mixture, $b \approx 0.63$~\cite{KA}. 
The values found above are somewhat larger, but it is hard to know
how preasymptotic effects influence the numerical data. 
Moreover, the value closest to $b$, $\mu \approx 0.73$, 
is obtained for $T=0.42$, a temperature already lower than 
the mode-coupling singularity located at $T_c \approx 0.435$ in this
system. MCT also provides a prediction for the height of the peak, 
$\chi_4^* \sim t^{*1/\gamma}$, where $\gamma$ was measured to be $\approx 2.3$, leading
to $\lambda=1/\gamma \approx 0.43$. This prediction is in good agreement with the
results of Ref.~\cite{WBG} where $\chi_4(t^*) \sim t^{*0.4}$ was reported.

If on the other hand one insists to use a non-cooperative 
kinetically constrained model 
to describe the Lennard-Jones liquid, the small value of the short 
time exponent $\mu$
forces one to choose a `fragile' KCM model,
such as the East model described above, where
the exponent for the dynamic susceptibility is found to be 
much smaller than the diffusive value $\mu=2$, and indeed to decrease 
when temperature is decreased.
However, the explanation given in Ref.~\cite{nef} that the large dynamic
length-scales observed in the Lennard-Jones system are due to the fact
that the underlying KCM model is relatively strong becomes hard to 
reconcile with the present 
results. This findings can therefore be added to the list 
of unusual features displayed by
supposedly `fragile' numerical 
models for super-cooled liquids~\cite{gilles}. 
On the other hand, note that our results do not discard the possibility that
cooperative KCM (in a proper density or temperature regime) display a four point correlation and susceptibility
quantitatively similar to the one of the Lennard-Jones liquid. Indeed, as stressed in e.g \cite{TBF}, 
for these models one expects a first regime of slowing down of dynamics due to an avoided mode-coupling transition. The
susceptibility and four point correlation could then well be quantitatively comparable to that of Lennard-Jones liquids.

\begin{figure}
\psfig{file=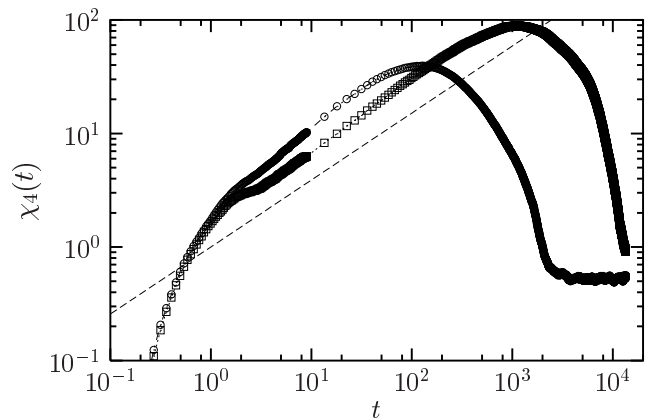,width=8.5cm}
\caption{\label{softsphere} Dynamic susceptibility $\chi_4(t)$
at $T=0.3$ and $0.26$ (from left to right)
in a log-log plot as a function of time for the soft-sphere 
binary mixture of Ref.~\cite{Grigera,Dave}. The data was kindly provided to us by 
D. Reichman and R. A. Denny. The straight line represent the MCT prediction for the power-law behaviour before the peak.}
\end{figure}

Finally, it is of course a natural question to ask whether the above agreement 
between MCT predictions and numerical results is only restricted to the Lennard-Jones 
system. Using the unpublished data of Ref.~\cite{Dave} for a soft-sphere 
binary mixture where 
$T_{c}\simeq 0.22-0.24$~\cite{Barrat,Grigera} 
we actually found very similar results.
Close to $T_{c}$ a power-law behaviour of $\chi_4$ as a function
of time can again be observed. For instance, 
$\chi_4 \sim t^{0.63}$ for $T=0.26$.
In Fig.~\ref{softsphere} we plot $\chi_4$, defined as 
in Ref.~\cite{FP}, as a function of time. We also display
the power-law behaviour predicted by MCT before the peak with the exponent $b \simeq 0.59$ taken
from Ref.~\cite{Barrat}. There is a similar agreement between the exponent $\lambda$ measured 
from the height of the peak and the value of $1/\gamma$ extracted from an MCT analysis of the data.

The fact that the predictions of MCT for the four-point susceptibility are in reasonable agreement 
with numerical simulations in both systems is significant, since the exponents $b$ and $1/\gamma$ 
are measured on (local) two point functions and $\mu$ and $\lambda$ on four-point functions. The relation between
these exponents test a rather deep structural prediction of MCT that relates time 
scales to length scales \cite{BB1}.
More numerical work, on other model systems with different values of $b$, for example, 
would be needed to establish
more firmly whether the coincidence observed in the present paper is or not accidental. 

\subsection{A growing length scale?} 

We focus now more directly on the dynamic lengthscale. 
In previous works, the dynamic lengthscale $\xi$ extracted 
from four-point correlations was measured either at fixed temperature
for various times $t$ where it was found to be 
non-monotonic~\cite{Glotzer,Glotzer2,pan} but monotonic in \cite{heuer}, 
or at fixed time $t = \tau(T)$, 
for different temperatures, where it is found to be 
increasing when the temperature decreases~\cite{Onuki,Glotzer,WBG}. 
In practice, to extract $\xi(t,T)$ from the four-point correlation function 
either in real space or
in Fourier space, one needs to postulate a specific 
functional form of $G_4$. 
In this respect, the results of the previous 
section on simple lattice KCM's with no underlying liquid
structure prove instructive. It is clear that 
with data similar to Fig.~\ref{3dfa2}, 
but obtained with much smaller system sizes, 
with much less statistics, and polluted by the underlying structure
of the liquid, the precise extraction of dynamical 
length-scales from Molecular Dynamics simulations
is not an easy task. More fundamentally, 
extracting $\xi$ from fitting either $G_4(r,t)$ 
or $S_4(k,t)$ to a time-independent scaling form necessarily
biases the data as discussed above.
This also shows that it is a much easier and safer procedure 
to work say at $t=\tau(T)$ and different temperature 
to observe the growth of a cooperative length 
$\xi(\tau,T)$ when decreasing $T$.
On the other hand, it is not {\it a priori} granted that 
the growth law of $\xi$ with $t = \tau(T)$
when changing $T$ is identical to that of $\xi(t,T)$ with $t$ at 
a given temperature $T$. We will not be able to answer this
question with our numerical data.

\begin{figure}
\psfig{file=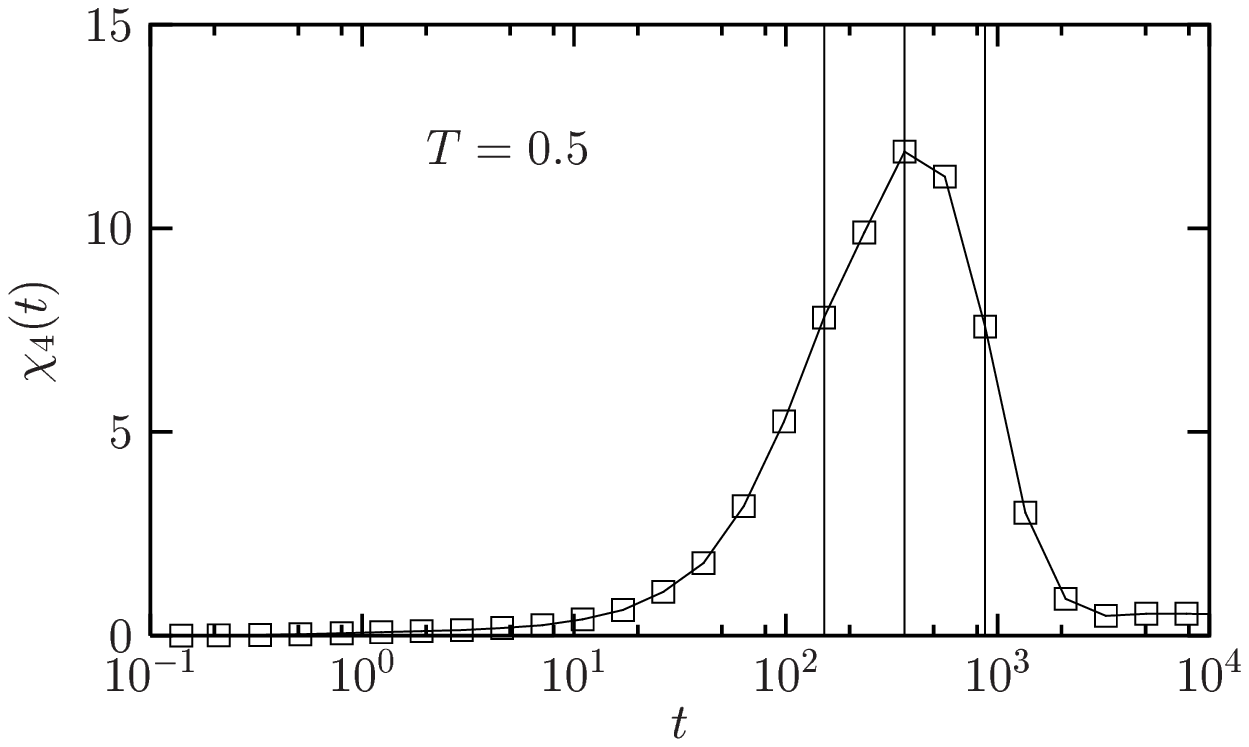,width=8.5cm}
\psfig{file=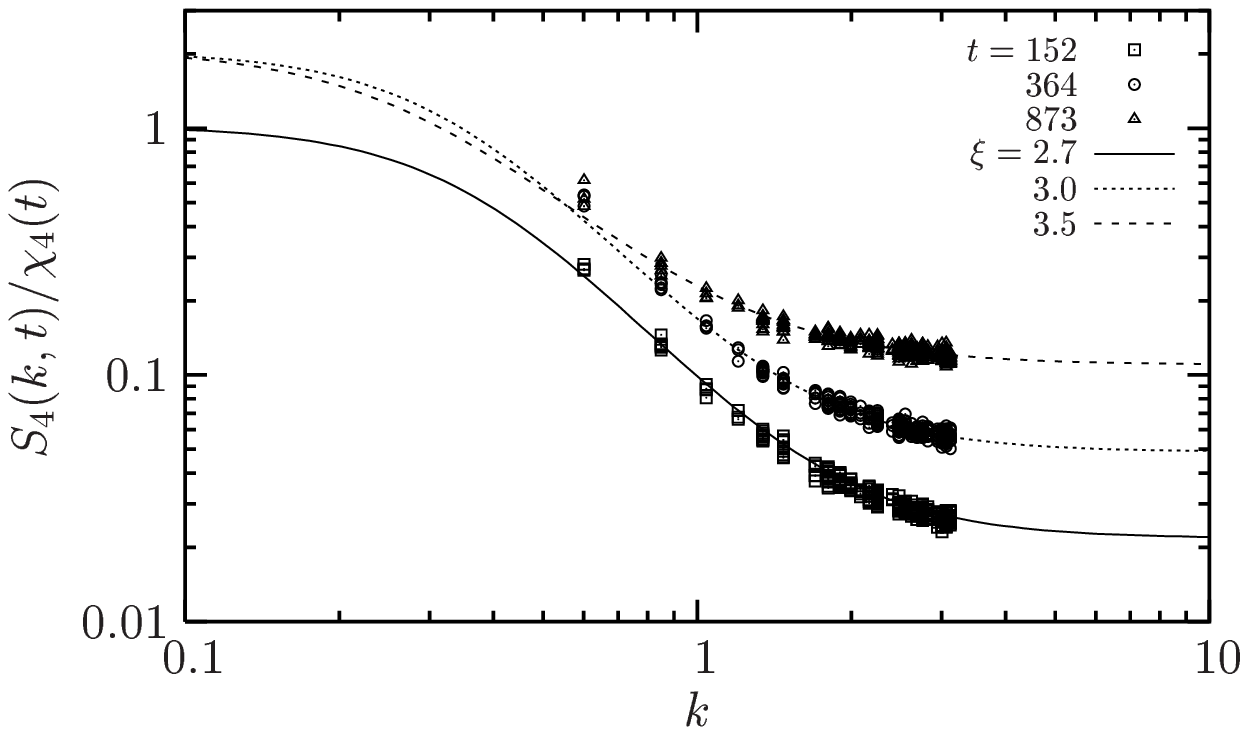,width=8.5cm}
\caption{\label{monolj} Left: dynamic susceptibility at $T=0.5$
and $q=4.21$. The vertical lines indicate the times
at which $S_4(k,t)$ is evaluated in the bottom 
figure. 
Right: the corresponding three $S_4(k,t)$ (the last two have been 
multiplied by 2 for clarity). Lines are fits to
the form (\ref{fifit}), the $k \to 0$ limit being 
fixed by the value of $\chi_4(t)$, with a monotonically 
growing length scale $\xi(t)$.}
\end{figure}

With the above caveats in mind, we present in 
Fig.~\ref{monolj} some numerical data 
in the binary Lennard-Jones mixture
at a fixed temperature, $T=0.5$, and three different times
which fall before, at and after the peak in $\chi_4(t)$. 
The difficulty of getting clear-cut quantitative determinations
for $\xi$ are obvious from Fig. \ref{monolj}. One would need  
much larger system sizes to properly measure $S_4(k,t)$
at small wavectors, large times and low temperatures. The system
simulated here contains $1372$ particles. One could possibly 
increase the number of particles by a factor 10, 
but the increase in linear size would be very modest, 
a factor $10^{1/3} \approx 2.15$.
Nonetheless, we have fitted the data in Fig.~\ref{monolj}
with a simple empirical form, 
\be
S_4(k,t) = \frac{\chi_4(t)-C}{1+ (k \xi)^\beta} + C,
\label{fifit} 
\ee 
for $0 \leq k < k_0$, $k_0 \approx 7.21$ 
being the position of the first peak in the static structure factor.  
As for the $d=3$ FA model, the exponent $\beta(t)$ and 
the dynamic length $\xi(t)$ are fitting parameters. There is 
an additional free parameter, the additive constant $C$ in Eq.~(\ref{fifit}), 
which accounts for the fact that the structure of the liquid 
starts to be visible and creates some signal in $S_4(k,t)$ when
$k \to k_0$. 
The results of the fitting procedure are presented in Fig.~\ref{monolj}
with lines going through the data. Note that the fits 
in Fig.~\ref{monolj} are constrained at low $k$
by the value of the dynamic susceptibility $\chi_4(t)$. 
The most important result from Fig.~\ref{monolj} is that if the functional
form of $S_4(k,t)$ is given some freedom, here via the time 
dependent exponent $\beta(t)$, the extracted dynamic
lengthscale $\xi(t)$ indeed continues to grow
monotonically after the peak of the dynamic susceptibility, contrary
to reported previously~\cite{Glotzer,Glotzer2,pan}, but in agreement with \cite{heuer}.
We emphasize once more that this result physically makes sense. 
At times much larger than $t^*$, only very rare but very large 
dynamical domains contribute to the dynamic structure factor, so that
spatial correlations are weak, but extremely long-ranged. The existence 
of an ever growing length scale is supported by any model with an hydrodynamical 
limit (such as the phonon or defect models studied here) and 
is in a sense trivial. The really interesting 
piece of information is the value of this length scale for $t=\tau_\alpha$,
i.e. when the relevant relaxation processes take place.

We conclude that our numerical data are not inconsistent with a 
monotonically growing length scale even for $t > \tau$, although 
addressing more quantitative issues such as 
functional form at the growth law and its temperature dependence would 
require quite an important, but certainly
worthwhile, numerical effort. 

\section{Conclusion and final comments}\label{conclusion}

Let us summarize the results and the various points made in this rather 
dense paper. 
First, we have computed numerically and analytically, exactly or
approximatively, the four-point correlation function designed to characterize
non trivial cooperative dynamics in glassy systems
within several theoretical models: mode-coupling theory, 
collectively rearranging regions, 
diffusing defects, kinetically constrained models, elastic/plastic
deformations. 
The conclusion is that the behaviour of $\chi_4(t)$ is rather rich, with 
different regimes summarized in the Introduction and in Fig. 1. We have 
computed the early time exponent $\mu$ and the peak exponent $\lambda$ for
quite a few different models of glass forming liquids, and shown that the 
values of these exponents resulting from these models are quite different,
suggesting that the detailed study of $\chi_4(t,T)$ should allow one to
eliminate or confirm some of the theoretical models for glass formation. 

In this spirit, we first simulated some non-cooperative KCMs as the 
one-spin facilitated FA model 
in $d=1$ and
$d=3$ and the East model. The assumption of 
point-like defects that diffuse, possibly with an anomalous diffusion exponent,
gives a good account of the shape of the four-point correlation function 
and of the four-point 
susceptibility which are in quantitative agreement with the above results for
the independent defects model. 
For strong glasses such as ${\rm S}{\rm i}{\rm O}_{2}$, 
where the relaxation is
due to defect diffusion, our results should be quantitative. It would be very
interesting to reconsider numerical simulations of the dynamics of ${\rm S}{\rm i}{\rm O}_{2}$
under the light of the present paper to check in more details that the defect picture
is indeed correct in this case (note that our results should 
enable one to extract, in 
principle, the properties, density and relaxation times of defects from the 
four-point correlation function). 
For the $d=3$ one-spin facilitated FA model, we see clear indications of the 
interactions between defects as time increases. 
This leads to small deviations of the
numerically obtained exponents with respect to those predicted by our analysis
of the independent defect model, which does not account for interactions
between defects.
As far as the identification of a growing length scale $\xi(t)$ from numerical data, 
we have seen that even within 
this simplified lattice models, this can be a rather difficult
task. Our results points toward a dynamical correlation length that
grows forever and a behaviour of $S_{4} (k,t)$ 
different from the Ornstein-Zernike form but with similar asymptotic
behaviour. We leave the study of cooperative KCM,
for which a more complicated behaviour should occur, for future work. In particular, the
detailed form of $S_{4} (k,t)$ should contain information about the inner structure of the
corresponding defects. 

We have also analyzed the four-point susceptibility of both a Lennard-Jones system and
a soft-sphere system, and shown that the initial 
exponent $\mu$ of the four-point susceptibility is decreasing with the temperature
and rather small, $\mu < 1$. We have found, perhaps unexpectedly, a reasonable agreement for
$\mu $ and $\lambda $ with the predictions of MCT but not with other theoretical scenarii, 
such as simple diffusing defects, strong KCMs or CRR (although this might be a question of 
temperature and time scales, since both CRR and cooperative KCMs are supposed to apply closer
to the glass transition temperature). Finally we confirm that the extraction of the growth law 
of $\xi(t)$ at a given temperature is difficult, and we can only say at this stage that the data is not 
incompatible with the idea that $\xi(t)$ grows monotonically, even beyond $t = \tau_\alpha$, 
in the Lennard-Jones system.

As for further work and perspectives, we think 
that the following points would be worth investigating. 
First, it would be very interesting to develop a detailed theory of 
the crossover
between the elastic regime described in Section~\ref{Elastic} and the Mode-Coupling 
$\beta$ relaxation regime. Is it possible, in particular, to describe approximately 
the `melting' of the glass as one approaches the Mode-Coupling transition 
temperature from below? Second,
we only considered systems in equilibrium. 
One in fact expects that the four-point susceptibility 
also contains
very useful information in the aging regime (see \cite{Leticia, mayer}). 
Detailed predictions
in this regime may enable one to probe the mechanisms for slow dynamics and the issue of the cooperative 
length at low temperature in the aging regime \cite{Leticia}. In particular, the elastic contribution should not
age whereas the CRR contribution (characterized by the same exponent $\mu$) should exhibit some aging, 
possibly allowing one to separate the two effects. Third, the quantitative study of four-point functions in 
cooperative KCMs where defects have a complex inner structure would be very interesting, since it is clear from the
present paper that simpler KCMs seem to fail at describing quantitatively $\chi_4(t)$ in fragile systems. Fourth, it
would be interesting to define more complicated correlation functions, for example, a fully general four point function,
or higher order correlation functions, in order to test in a more stringent way the idea 
of cooperativity in glassy 
systems, and distinguish systems where the growth of $\chi_4(t)$ is trivial, such as elastic solids, from 
those in which 
a truly non trivial cooperativity governs the dynamics. Finally, it seems clear that this issue of 
cooperativity and
its associated length scale can only be convincingly settled if long time scales, low temperature 
regimes can be
probed quantitatively in experimental systems. We hope that the present paper will motivate ways to 
directly 
access four-point functions experimentally in glassy systems (see \cite{mayer}); natural candidates 
for this are 
colloids \cite{Weeks} and granular materials \cite{Dauchot,Alex}, although there might be ways to 
investigate this 
question in molecular glasses and spin-glasses as well \cite{ustocome}.

\begin{acknowledgments}
Fig.~\ref{softsphere} was obtained from the unpublished data 
of D.R. Reichman and R.A. Denny. We are very grateful to them for providing these results. 
We thank E. Bertin, O. Dauchot, J.P. Garrahan and D.R. Reichman for discussions. 
G. B. is partially
supported by the European Community's Human Potential Programme
contracts HPRN-CT-2002-00307 (DYGLAGEMEM).
C.T. is supported by the European Community's Human Potential Programme
contracts HPRN-CT-2002-00319 (STIPCO).
\end{acknowledgments}

\section*{Appendix A: Dynamics of elastic networks}

\subsection*{A.1 The four-point correlation function - Over-damped case}

We will define $G_4(\vec r,t)$ for the elastic model defined in the text as: 
\be
G_4(\vec r,t)=\langle \cos q \left[\phi(\vec r,t)-\phi(\vec r,0)\right]\cos q 
\left[\phi(\vec r=0,t)-\phi(\vec r=0,0)\right]
\rangle-C^2(q,t),
\ee
which is equivalent to:
\ba\nonumber
G_4(\vec r,t)&=&\frac12 \langle \cos q \left[\phi(\vec r,t)-\phi(\vec r,0)+\phi(\vec r=0,t)-\phi(\vec r=0,0)
\right]
\\\nonumber
&+& \frac12 \langle \cos q \left[\phi(\vec r,t)-\phi(\vec r,0)-\phi(\vec r=0,t)+\phi(\vec r=0,0)
\right]-C^2(q,t).
\ea
Using the fact that the field $\phi$ is Gaussian, we finally find:
\be
G_4(\vec r,t)=C^2(q,t)\left(\cosh(2q^2 R(\vec r,t))-1\right),
\ee
where:
\ba\nonumber
R(\vec r,t)&=& \langle(\phi(\vec r,t)-\phi(\vec r,0))(\phi(\vec r=0,t)-\phi(\vec r=0,0))\rangle\\
&=&\frac{T}{\kappa} \int \frac{d^dk}{(2 \pi)^d k^2} e^{-i\vec k \cdot \vec r}(1-e^{-\kappa k^2 t})
\ea
Hence, 
\be
R(\vec r,t)=\frac{T}{\kappa} (\kappa t)^{1-d/2} F(\frac{r}{\sqrt{\kappa t}})
\ee
with:
\be
F(z)=z^{2-d} [I(\infty)-I(z)]; \qquad I(z)=\int \frac{d^dw}{(2\pi)^d w^2} e^{-iw_1-\frac{w^2}{z^2}}.
\ee
We thus see immediately that $G_4(\vec r,t)$ will be governed by a 'diffusive' correlation length 
$\xi(t) \sim \sqrt{\kappa t}$, as expected from the structure of the Langevin equation that describes
relaxational dynamics. Note that for under-damped dynamics, sound waves would change this scaling.

It is useful to consider the following quantity:
\be 
J(z)=\frac{\partial I(z)}{\partial(\frac{1}{z^2})}=\int \frac{d^dw}{(2\pi)^d} e^{-iw_1-\frac{w^2}{z^2}}.
\ee
In $d=3$, after integrating over $dw_1$, one has
\be
J(z)= \frac{1}{8 \pi^{3/2}} z^3 e^{-\frac{z^2}{4}}
\ee
and:
\be
I(z)=\frac{1}{4 \pi^{3/2}} \int_z^\infty e^{-\frac{u^2}{4}} du
\ee
Therefore, for $z \ll 1$, one finds $F(z) \simeq (4\pi z)^{-1}$ and $R(\vec r,t) \simeq T/(4\pi \kappa r)$, 
whereas for $z \gg 1$, $F(z) \simeq (2 \pi^{3/2})^{-1} \exp(-z^2/4)/z^2$. 
Thus, for $r \ll \xi(t)$ and $\kappa \Lambda^2 t \gg 1$, the four-point correlation function behaves as:
\be
G_4(\vec r,t)=f_q^2 \left(\cosh(\frac{T q^2}{2\pi \kappa r})-1\right).
\ee

\subsection*{A.2 The four-point correlation function - Under-damped case}

We have now:
\be
m\frac{\partial^2 \phi(\vec r,t)}{\partial^2 t}= \kappa \Delta \phi(\vec r,t)
\ee
which has for solutions in the Fourier space:
\be
\phi_k(t)=\exp( ik Vt) \phi_k(0),
\ee
with $V=(\kappa/m)^{1/2}$.
We now have:
\be
\label{beta}
\langle [(\phi(\vec r,t)-\phi(\vec r,0)]^2\rangle= 
\frac{2T}{\kappa} \int \frac{| \exp(ik Vt)-1|^2}{k^2} \frac{d^dk}{(2\pi)^d}= 
\frac{4T}{\kappa} \int (1-\cos[Vkt])dk.
\ee
In $d=3$, we find obviously the same result for $f_q$ and $G_4$ as above, but $R(\vec r,t)$ is now equal to:
\be
R(\vec r,t)=\frac{T}{\kappa} \int \frac{d^dk}{(2 \pi)^d k^2} e^{-i\vec k \cdot \vec r}(1-\cos(k Vt))
\ee
that we write:
\be
R(\vec r,t)=\frac{T}{\kappa} [I(\vec r,0)-I(\vec r,t)]
\ee
where:
\be
I(\vec r,t)=\int \frac{d^dk}{(2 \pi)^d k^2} e^{-i\vec k \cdot \vec r}\cos(k Vt)
\ee
By introducing $z=Vt/r$ and changing variable $q\equiv r k$, and also $u=\cos\theta$ and 
integrating over $u$ one finds:
\be
I(\vec r,t)= \frac{2\pi}{r} \int dq q^{-1} [\sin(q(1+z))+\sin(q(1-z))]
\ee
Consider the first term:
\be
I(\vec r,t)= \frac{2\pi}{r} \int dq q^{-1} \sin(q(1+z))
\ee
Changing variable $v=q(1+z)$ directly shows that this integral do not depend on $z$, as long as $(1+z)$ is positive. 
This is true for the other integral, which does not depend on $z$ as long as $1-z$ is positive. If  $1-z$ is negative 
then the integral changes sign. Therefore we have that $I(\vec r,t)=I(\vec r,0)$ if $z<1$, and $I(\vec r,t)=0$ if $z>1$. 
Therefore $R(\vec r,t)=0$ if $z<1$ and  $R(\vec r,t)=\frac{T}{4\pi\kappa r}$ when $z>1$. The result is very intuitive: 
when $z<1$ the information does not have time to travel the distance $r$ and there are no correlation. For $z>1$ the 
two regions are ``connected'' and one finds the free field correlations.  Brownian and Newtownian dynamics 
furnish the same correlation for a given $r$ when the time diverges, as we expect. Finally, it is straightforward
to obtain the result quoted in the text for $\chi_4(t)$.

\subsection*{A.3 The low dimensional case}

We give here, without much details, the results for elastic networks in $d=1$ and $d=2$. In
$d=1$, as is well known, each particle wanders arbitrary far from its initial position but in
an anomalous, sub-diffusing way, as $t^{1/4}$. Correspondingly, the dynamical structure factor 
decays as a stretched exponential:
\be
\ln C(q,t) \sim \frac{T}{\kappa} q^2 t^{1/2}
\ee
Note that the $t^{1/4}$ comes from a collective displacement of the cages, and is similar to the 
anomalous diffusion observed for hard spheres in one dimensions, since the latter problem can be
mapped onto the Edwards-Wilkinson problem i one dimension \cite{Arrantia,Alex}. 
We expect that the results obtained here for $G_4$ should also hold for this case as well. In fact,
this model was recently discussed in the context of a simple $d=1$ granular compaction model, see \cite{Alex}.

In $d=2$, the displacement grows logarithmically with time, leading to a power-law decay of the
dynamical structure factor with a $q$ dependent exponent:
\be
C(q,t) \sim t^{-y} \qquad y = \frac{q^2 T}{8 \pi \kappa}.
\ee
Turning now to $\chi_4(t)$, we find that after a short transient, $\chi_4(t)$ grows as $t^{1/2}$ 
in $d=1$ and behaves as $t^{1-2y}$ in $d=2$.

\section*{Appendix B: Calculations for the defect model}

In section \ref{defect} we have reduced the computation of $G_4(r,t)$ and $\chi_4(t)$
to probability distributions of a single random walk. In the following we shall show 
how these quantities can be computed in any spatial dimension.

Let us call $F^z_x(u)$ be the probability that a random walk starting in $z$ reaches $x$
for the first time at time $u$. $P^z_x(t)$, the probability that a vacancy 
starts in $z$ at time zero and reaches for the first time $x$ at a time less than $t$, reads:

\begin{equation}
P^z_x(t)=\int_0^t F^z_x(u) du
\end{equation}

Therefore, we need to calculate $F^z_x(u)$. The trick to do that is writing a linear equation relating 
$F^z_x$, that we want to compute, to $P^z(x,t)$, the probability that a random walk with self
diffusion coefficient $D$, starting in $z$, is in x at time t, which is well known.
This linear equation is:
\begin{equation}
P^z(x,t)=\delta_{x,z}\delta_{t,0}+\int_0^t F^z_x(u)P^x(x,t-u) du\quad.
\end{equation}
By  taking Laplace transform (from now on $s$ is the variable
conjugated to $t$ and $\cal{L}$ indicates the Laplace transform) we obtain:

\begin{equation}
\label{10}
F^z_x(t)={\cal{L}}^{-1}\left(\frac{{\cal{L}} P^z(x,s)-\delta_{x,z}}{{\cal{L}} P^x(x,s)}\right)(t)
\end{equation}

and
\begin{eqnarray}
\label{10b}
P^z_x(t)&&=\int_0^t {\cal{L}}^{-1}\left(\frac{{\cal{L}} P^z(x,s)-\delta_{x,z}}{{\cal{L}} P^x(x,s)}\right)(t') 
dt'\\
&&= {\cal{L}}^{-1}\frac{1}{s}\frac{{\cal{L}} P^z(x,s)-
\delta_{x,z}}{{\cal{L}} P^x(x,s)}
\end{eqnarray}

A similar strategy can be used to calculate  $P^z_{x, {\overline y}}(t)$.
Indeed the following equality hold

\begin{eqnarray}
P^z_{x, {\overline y}}(t)&=&\int_0^t F^z_{{x,\overline y}}(t')P^x_{{\overline y}}(t-t') dt'\nonumber\\
P^z_{y, {\overline x}}(t)&=&\int_0^t F^z_{{\overline x},y}(t')P^y_{{\overline x}}(t-t') dt'\nonumber\\
\end{eqnarray}
where $F^z_{{\overline x},y}(t)$ is the probability that a random walk starting in $z$
at time zero reaches $y$ for the first time at t but never touches $x$ at
$s\leq t$. 
Therefore, in order to calculate 
$P^z_{x, {\overline y}}(t)+P^z_{y, {\overline x}}(t)$
we need to calculate $F^z_{x, {\overline y}}(t)+F^z_{y, {\overline x}}(t)$. 
It is immediate to check that the following equations hold for any choice of $x,z,y$

\begin{equation}
\left\{
\begin{array}{ll}
F^z_x(t) & =\delta_{x,z}\delta_{t,0}+\int_{0}^t ds F^z_{{\overline x},y}(s)F^y_x(t-s)
+F^z_{x,{\overline y}}(t)\\
F^z_y(t) & =\delta_{y,z}\delta_{t,0}+\int_{0}^t ds F^z_{x,{\overline y}}(s)F^x_y(t-s)
+F^z_{{\overline x},y}(t)
\end{array}
\right.
\end{equation}

which implies, again by Laplace transform ($z$ is always different
from $x$ and $y$ in the following so we will skip the Kronecker deltas),

\begin{equation}
\label{12}
F^z_{{\overline x},y}(t)+F^z_{x,{\overline y}}(t)={\cal{L}} ^{-1}\frac{{\cal{L}}
F^z_x(s)+{\cal{L}}F^z_y(s)}{{\cal{L}}F^x_y(s)+{\cal{L}}F^{x}_x(s)}
\end{equation}

Using the expression (\ref{10}) for the $F^x_y(s)$ we get

\begin{equation}
F^z_{{\overline x}}(y,t)+F^z_{{\overline y}}(x,t)={\cal{L}} ^{-1}\frac{{\cal{L}}
P^z(x,s)+{\cal{L}}P^z(y,s)}{{\cal{L}}P^x (y,s)+{\cal{L}}P^{x} (x,s)}
\end{equation}

Furthermore $ P^y_{{\overline x}}(t)=1-P^{y}_{x} (t)$. Hence we obtain
\begin{equation}\label{}
{\cal{L}} P^y_{{\overline x}}(s)=\frac{1}{s} -{\cal{L}}P^{y}_{x} (s) =
\frac{1}{s} (1-\frac{{\cal{L}}P^{y} (x,s)}{{\cal{L}}P^{x} (x,s)})
\end{equation}

Finally, we obtain the expression for 
\begin{equation}
{\cal {L}} (P^z_{x, {\overline y}}(s)+P^z_{y,{\overline x}}(s))=\frac{{\cal{L}}
  P^z(x,s)+{\cal{L}}P^z(y,s)}{{\cal{L}}P^x(y,s)+{\cal{L}}P^{x} (x,s)}
\frac{1}{s} \left(1-\frac{{\cal{L}}P^{y} (x,s)}{{\cal{L}}P^{x} (x,s)}\right)
\end{equation}

An useful way to rewrite this expression is obtained 
by summing and subtracting the Laplace transform of $P^{z}_{x} (t)+P^z_y(t)$:
\begin{equation}
P^z_{x, {\overline y}}(t)+P^z_{y,{\overline x}}(t)=P^{z}_{x} (t)+P^{z}_{y} (t)-2{\cal{L}}^{-1}\frac{{\cal{L}}
  P^z(x,s)+{\cal{L}}P^z(y,s)}{{\cal{L}}P^x(y,s)+{\cal{L}}P^{x} (x,s)}
\frac{1}{s} \frac{{\cal{L}}P^{y} (x,s)}{{\cal{L}}P^{x} (x,s)}
\end{equation}

Finally putting together all the different terms we have:

\begin{equation}\label{}
\langle n_x(t)n_y(t)\rangle =\exp \left(-2\rho_{v}-2\rho_{v}N (t)+2\rho_{v}P^{y}_{x}
(t)+\rho_{v}G(t,x-y)
\right)
\end{equation}

where $N (t)=\sum_{z\neq x} P^z_x (t)$ is the average number of distinct sites
(minus 1) visited by a random walk during the interval of time $t$ and

\begin{equation}
G (t,x-y)={\cal{L}}^{-1}
\left[\sum_{z\neq x,y}
\frac{{\cal{L}}
  P^z(x,s)+{\cal{L}}P^z(y,s)}{{\cal{L}}P^x(y,s)+{\cal{L}}P^{x} (x,s)}
\frac{1}{s} \frac{{\cal{L}}P^{y} (x,s)}{{\cal{L}}P^{x} (x,s)}\right]
\label{G}
\end{equation}

Since 
\begin{equation}\label{}
\langle n_x(t)\rangle^{2}=\exp \left(-2\rho_{v}-2\rho_{v}N (t))
\right)
\end{equation}

the expression of 
$G_4$ is
\begin{equation}\label{final}
G_4 (x-y,t)=\exp \left(-2\rho_{v}-2\rho_{v}N (t)
\right)\left[\exp \left(2\rho_{v}P^{y}_{x}
(t)+\rho_{v}G (t,x-y)
\right)-1 \right]
\end{equation}
In the following we shall analyze separately the one dimensional
case, the three or higher dimensional case and the two dimensional
case.

\section*{B.1 One Dimension}
\label{1D}

Consider a symmetric random walk on a one dimensional lattice with
lattice spacing $a$, by Laplace
transforming the master equation 

\begin{equation}
\frac{dP^z(x,t)}{dt}=\frac{P^z(x+a,t)+P^z(x-a,t)-2P^z(x,t)}{2}
\end{equation}

one immediately obtains
%In the continuum limit 
%the evolution equation for the

\begin{equation}
{\cal{L}}P^z(x,s)=\int_{-\pi/a}^{\pi/a}\frac{dk}{2\pi}\frac{e^{ik (x-z)}}{\zeta(k)+s}
\end{equation}

where $\zeta(k)=(1-\cos k)$.
% and the between nearest neigbor on
%the lattice is taken equal to one.
In the continuum limit $a\to 0$, $(x-y)\propto a$ $\sqrt{Dt/2}\propto a^2$,
the above integral can be solved with the well known result

\begin{equation}
{\cal{L}} P^z(x,s)=\frac{1}{ \sqrt{4 D s}}e^{\frac{-\sqrt{s} |x-z|}{\sqrt D}}
\end{equation}

which correspond to the solution of the diffusion equation for a one
dimensional Brownian
motion with diffusion coefficient $D$, i.e.

\begin{equation}
\frac{dP}{dt}=D\frac{d^2P}{dx^2}
\end{equation} 

Let us now compute all the functions needed to get $G_{4}$.

First

\[
N (t)=\sum_{z\neq x} P^z_x(t)=\sum_{z\neq x}{\cal{L}}^{-1} \left(\frac{1}{s} 
\frac{{\cal{L}}P^{z} (x,s)}{{\cal{L}}P^{x} (x,s)}\right) (t)
\]

where we used equation (\ref{10b}).
When $t>>1$ we get 

\[
N (t)=4\frac{\sqrt{Dt}}{\sqrt{\pi}}
\]

Second, 
using the expression (\ref{G}) of $G$ in terms of ${\cal{L}} P^z(x,s)$ we get:

\[
{\cal{L}}G (s,x-y)=2\frac{\sqrt{D}}{s^{3/2}}\frac{e^{\frac{-\sqrt{s} 
|x-y|}{\sqrt D}}}{e^{\frac{-\sqrt{s} |x-y|}{\sqrt D}}+1}
\]

Changing variable in the Inverse Laplace transform we get:

\[
G (t)=4\sqrt{Dt}f \left( \frac{|x-y|}{\sqrt{2Dt}}\right)
\]

where $f \left(\frac{|x-y|}{\sqrt{2Dt}}\right)$ equals 

\[
f \left(\frac{|x-y|}{\sqrt{2Dt}}\right)=\int_{-i\infty -\gamma}^{+i\infty -\gamma}
\frac{e^{\frac{-\sqrt{2s} |x-z|}{\sqrt Dt}}}{e^{\frac{-\sqrt{2s} |x-z|}{\sqrt Dt}}+1}
e^{-s}\frac{ds}{s^{3/2}}\]

Finally $P^{y}_{x} (t)$ can be computed easily but it is always much
smaller than the other terms in the exponential so we are going to
neglect it. The resulting expression for $G_{4}$ is:

\begin{equation}\label{g41d}
G_{4} (x-y,t)=\exp \left(-2\rho_{v}-\frac{8\rho_{v}}{\sqrt{\pi}}\sqrt{Dt}
\right)\left[\exp \left(\rho_{v}2\sqrt{Dt}f\left( \frac{|x-y|}{\sqrt{2Dt}}\right)
\right)-1 \right]
\end{equation}

Note that the typical time-scale is $\tau=\frac{1}{\rho_{v}^{2}D}$ and
since we focus on $\rho_{v}\rightarrow 0$ we can rewrite the above
expression as:

\begin{equation}\label{g41d2}
G_{4} (x-y,t)=\exp \left(-\frac{\sqrt{8}}{\sqrt{\pi}}\sqrt{t/\tau }
\right)\left[\exp \left(2\sqrt{t/\tau }f\left(
\rho_{v}\frac{|x-y|}{\sqrt{2t/\tau }}\right)
\right)-1 \right]
\end{equation}

Integrating over $x-y$ to get the $\chi_{4}$ we find:

\begin{equation}\label{chi41d}
\chi_{4} (t)=\frac{2}{\rho_{v}}\exp \left(-\frac{8}{\sqrt{\pi}}\sqrt{t/\tau }
\right)\sqrt{2t/\tau }\int_{0}^{+\infty}dx \left[\exp \left(2\sqrt{t/\tau }f(x)
\right)-1 \right]
\end{equation}

In particular when $t/\tau \ll 1$ we have 

\begin{equation}\label{chi41da}
\chi_{4} (t)\propto \frac{1}{\rho_{v}} (t/\tau )
\end{equation}

The interpretation of this result is that at short times the defects
do not intersect and the $\chi_{4}$ is just the square of the number 
of average sites visited by a random walk until time $t$. We'll see
that this interpretation is indeed correct in any dimension.

Finally, after some algebra it is possible to obtain from 
(\ref{chi41da}) that $\chi_{4} (t)\simeq\frac{c}{\rho_v} \exp \left(-\frac{4}{\sqrt{\pi}}\sqrt{t/\tau }
\right)$ at very large times ($c$ is a numerical constant). Thus,
as found in simulations, the normalized $\chi_{4}$ does not go to zero as it happens in three dimensions.

\section*{B.2 Three dimension and higher}\label{3D}

Consider a symmetric random walk on a cubic lattice, the  general expression for $P^z(x,s)$ is

\begin{equation}
P^z(x,s)=\int_{BZ}\frac{d^{d}k}{(2\pi)^{d}}\frac{e^{ik (x-z)}}{\zeta(k)+s}
\end{equation}
where $BZ$ means Brillouin zone and
$\zeta(k)=\sum_{i=1}^{d}(1-\cos k_{i})$ for a hyper-cubic lattice ($k_i$ is the component of $\vec{k}$ in the direction $i$).
Also in this case we consider the  continuum limit 
$(x-y)/\sqrt{Dt/2}\propto O (1)$ and look for times $t$ much larger than one.

Let us again compute all the needed quantities.
First, $N (t)$. In this case for $t>>1$ we find that 
\[
N (t)= D\left(\int_{BZ}\frac{d^{d}k}{\pi \zeta(k)}\right)^{-1}{\cal{L}}^{-1} \frac{1}{s^{2}}
\]

Hence $N (t)= c_1tD$
where $c_1=\left(\int_{BZ}\frac{d^dk}{\pi \zeta(k)}\right)^{-1}$.

Again, we neglect the $P^{y}_{x} (t)$ term and we focus on $G$ in the
continuum limit, for $t\gg a$ we get:

\[
{\cal{L}}G=\frac{1}{s^{2}}\frac{\int_{BZ}\frac{d^{d}k}{(2\pi)^{d}}\frac{e^{ik (x-z)}}
{Dk^{2}+s}}{\left( \int_{BZ}\frac{d^{d}k}{(2\pi)^{d}}\frac{1}{Dk^2}\right)^{2}}
\]

Changing variable in the Inverse Laplace Transform we get:

\[
G (t)=D^2\int_{-i\infty -\gamma}^{+i\infty -\gamma}e^{ts}\int_{BZ}
\frac{d^{d}k}{(2\pi)^{d}}\frac{e^{ik (x-z)}}{Dk^{2}+s} \frac{\exp (ts)}{C^{2}s^{2}}ds
\]

%(where $\gamma$
% is chosen as usual 
%in order that the contour parallel to the $y$ axis in the first
% integral leaves all the poles of the integrand to the left)
Since we know the Inverse Laplace Transform of the function resulting
from the integral over $k$ (it is simply $P^{y} (x,t)$) and each $1/s$
adds an integral we finally get 
\[
G (t)= c_2(Dt)^{2}\int_{0}^{1}du\int_{0}^{u}dv \frac{e^{-\frac{(x-y)^{2}}{2Dtv}}}{(2Dtv)^{d/2}}
\]
where $c_2$ is a numerical constant of order unity. From this expression, we finally obtain 
\begin{equation}\label{g43d}
G_{4} (x-y,t)=\exp (-2\rho_{v}-2\rho_{v}c_1D t)
\left[ \exp \left(\rho_{v}(c_2Dt)^{2}\int_{0}^{1}du\int_{0}^{u}dv \frac{e^{-\frac{(x-y)^{2}}{2Dtv}}}
{(2Dtv)^{3/2}} \right)-1\right]
\end{equation}
and the results quoted in the main text. 

\section*{B.3 Two dimension}\label{2D}

In two dimension things are bit tricky because of log
corrections. Briefly, we obtain that 
\begin{equation}\label{g42d}
G_{4} (x-y,t)=\exp \left(-2\frac{c_3 t}{\tau \ln t}\right)
\frac{1}{\rho_{v}}c_4^{2}(t/\tau)^{2}\frac{1}{(\ln t D)^{2}}\int_{0}^{1}du\int_{0}^{u}dv
\frac{e^{-\frac{(x-y)^{2}}{2Dvt}}}{(2Dvt)}
\end{equation}
with $c_3$ and $c_4$ constants of order unity. 
Hence, integrating over $x-y$, we get:
\begin{equation}\label{chi42d}
\chi_{4}(t)=\exp \left(-2\frac{c_3 t}{\tau \ln t}\right)
\frac{1}{2\rho_{v}}c_4^{2}(t/\tau)^{2}\frac{1}{(\ln tD)^{2}}
\end{equation}

In particular when $t/\tau<<1$ we have 

\begin{equation}\label{chi41db}
\chi_{4} (t)\propto \frac{1}{\rho_{v}} \left(\frac{t}{\tau \ln t}\right)^2
\end{equation}
Again, since the number of sites visited in average by a RW in 2D goes like
$t/\ln t$, at short times $\chi_4$ is the square of the number of average
sites visited until time $t$.

\section*{B.4 Density--density correlations}

We now sketch the calculation for the density four point correlation, defined as: 
\begin{equation}
G_4^d(x-y,t)\equiv
\langle (\eta_x(t)\eta_x(0)-\rho^{2}) (\eta_{y} (t)\eta_y(0)-\rho^{2})\rangle-\langle\eta_x(t)\eta_x(0)\rangle_{c}^2
\end{equation}
with $\eta_x(t)=0,1$ if 
the site $x$ is empty or occupied at time $t$, respectively. We start from:
\begin{equation}
\langle\eta_x(t)\eta_x(0)\rangle_{c}^2=\left(\left[\frac{1}{V}\sum_{z, z\neq x}
(1-P^z(x,t))\right]^{N_v}-\rho^{2}\right)^{2}\nonumber
\end{equation}
Using that $\sum_{z} P^z(x,t)=1$ we get
\begin{equation}
\langle\eta_x(t)\eta_x(0)\rangle_{c}^2=\exp (-4\rho_{v}) \left(\exp (\rho_{v}P^{x} (x,t))-1 \right)^{2}
\end{equation}
In the limit $\rho_{v}\rightarrow 0$ we have:
\begin{equation}
\langle\eta_x(t)\eta_x(0)\rangle_{c}^2= \left( \rho_{v}P^{x} (x,t)) \right)^{2}
\end{equation}
Similarly we find that
\begin{eqnarray}
\langle\eta_x(t)\eta_x(0)\eta_y(t)\eta_y(0)\rangle&=&(\frac{1}{V}\sum_{z, z\neq x,y}
(1-P^z(x,t)-P^z(y,t)))^{N_v}\nonumber\\ 
&=& \exp \left(-4\rho_{v}+2\rho_{v}P^{x} (x,t)+2\rho_{v}P^{y} (x,t)  \right)\nonumber
\end{eqnarray}
Collecting all the pieces together we finally get at leading order in
$\rho_{v}$
\begin{equation}
G_4^d(x-y,t)=2\rho_{v}P^{y} (x,t)
\end{equation}
for $x\neq y$.
The interpretation of this eq is that the dynamical correlation
between $x$ and $y$ is due to the fact that the {\it same vacancy}
was in $x$ at time $0$ and $t$ at time $t$ or vice-versa.
Integrating over $x-y$ one finds that at long times 
$\chi_4(t)\propto 1/t^{d/2}$, 
showing no interesting structure.


\begin{thebibliography}{99}
\bibitem{foot} Note however that in a numerical simulation with a box of size 
say $L=10 \, a$, the elastic contribution will saturate after a time $L/V \sim 1$ ps. 
This contribution will therefore be hard to see unless much larger sizes 
can be simulated.
\end{thebibliography}
\end{document}